\documentclass[twocolumn,showpacs,preprintnumbers,amsmath,amssymb]{revtex4}
\usepackage{graphicx}
\usepackage{dcolumn}
\usepackage{bm}

\begin{document}
\title{Theoretical and simulation studies of characteristics of a Compton
light source} 
\author{C.~Sun\footnote{Currently at Lawrence Berkeley National Laboratory.}}
   \email{suncc@fel.duke.edu, CCSun@lbl.gov }
\author{Y.~K.~Wu}
\affiliation{Department of Physics, Duke University, Durham, NC 27708-0305, USA
and 
\\DFELL, Triangle Universities Nuclear Laboratory, Durham, NC 27708-0308, USA}
\date{\today}

\begin{abstract}

Compton scattering of a laser beam with a relativistic electron beam has been
used to generate intense, highly polarized and nearly monoenergetic x-ray or
gamma-ray beams at many facilities. 
The ability to predict the spatial, spectral and temporal characteristics of
a Compton gamma-ray beam is crucial for the optimization of the operation of a
Compton light source as well as for the applications utilizing the Compton beam.
In this paper, we present two approaches, one based upon analytical calculations
and the other based upon Monte
Carlo simulations, to study the Compton scattering process for various electron
and laser beam parameters as well as different gamma-beam collimation
conditions. These approaches have been successfully applied to characterize
Compton
gamma-ray beams, after being benchmarked against experimental results at the
High Intensity Gamma-ray Source (HI$\gamma$S) facility at Duke University.
\end{abstract}
\maketitle
\section{Introduction}

Compton scattering of a laser beam with a relativistic electron beam has been
successfully used to generate intense, highly polarized and nearly
monoenergetic x-ray or gamma-ray beams with a tunable energy at many 
facilities~\cite{H.R.Weller,Nakano:2001xp,2003SPIE.5197..241K}. 
These unique Compton photon beams have been used in a wide range of basic and
application research fields from nuclear physics to astrophysics, from medical
research to homeland security and industrial applications~\cite{H.R.Weller}. 

The ability to predict the spectral, spatial and temporal characteristics of
a Compton gamma-ray beam is crucial for the optimization of the gamma-ray beam
production as well as for research applications utilizing the beam. While the
theory of
particle-particle (or electron-photon) Compton scattering, which is equivalent
to the scattering between a monoenergetic
electron beam and a monoenergetic laser beam with zero transverse sizes, is
well documented in literature~\cite{QED_landau,CFT_landau,Jackson}, there
remains a need to fully understand the characteristics of the gamma-ray beam
produced by Compton scattering of a laser beam and an electron beam with
specific spatial and energy distributions, i.e., the beam-beam scattering. 

Study of beam-beam Compton scattering has been recently reported in
~\cite{Hartemann:2005zz,Brown:2004zz}. However, the algorithms used in
these works are based upon the Thomson scattering cross section, i.e., an
elastic scattering of electromagnetic radiation by a charged particle without the
recoil effect. 
For scattering of a high energy electron beam and a laser beam, the recoil of the
electron must be taken into account. 
The Compton scattering cross section has been used to study characteristics of
Compton gamma-ray beams by Duke scientists in
$1990$'s~\cite{Vladimir,Park_thesis}. 
However, the effects
of incoming beam parameters, and the effects of gamma-beam collimation were not
fully taken into account.

In this paper, we present two different methods, a semi-analytical
calculation and a Monte Carlo simulation, to study the Compton scattering
process of a polarized (or unpolarized) laser beam with an unpolarized electron
beam in the linear Compton scattering regime. 
Using these two methods, we are able to characterize a Compton 
gamma-ray beam with various laser and electron beam parameters,
arbitrary collision angles, and different gamma-beam collimation conditions. 

This paper is organized as follows. In Section II, we first review the calculation 
of the Compton
scattered photon energy for an arbitrary collision angle, and then introduce the
scattering cross section in a Lorentz invariant form. Based upon this cross
section, the spatial and spectral distributions as well as the polarization of a
Compton gamma-ray beam are investigated in particle-particle scattering cases.
In Section III, we discuss the beam-beam Compton scattering by considering
effects of the incoming beam parameters as well as the effect of the gamma-ray beam
collimation. Two methods, a semi-analytical calculation and a Monte Carlo
simulation, are then presented. Based upon the algorithms of these methods, two
computing codes, a numerical integration code and a Monte Carlo simulation
code, have been developed at Duke University. 
The benchmarking results and applications of these two codes are
presented in Section IV. The summary is given in Section V.

\section{Particle-particle scattering}
\subsection{Scattered photon energy }

\begin{figure}
\centering
\includegraphics[width=\columnwidth]{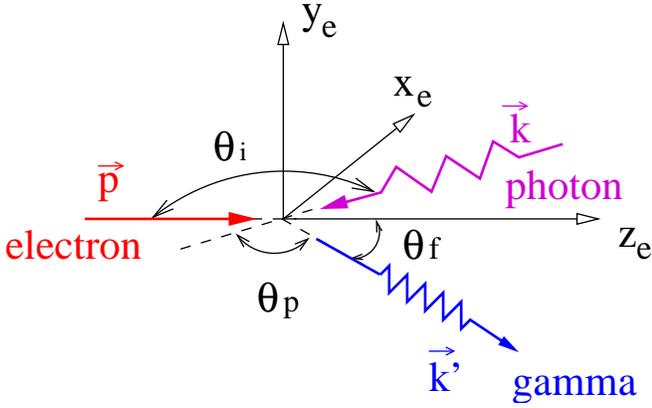}
\caption{\label{electron_lab_frame} Geometry of Compton scattering of an
electron and a photon in a lab frame coordinate system $(x_e,y_e,z_e)$ in which
the electron is incident along the $z_e$-direction.
The incident photon is propagating along
the direction given by the polar angle $\theta_i$ and azimuthal angle $\phi_i$.
The collision occurs at the origin of the coordinate system. After the
scattering, the scattered photon propagates in the
direction given by the polar angle $\theta_f$ and azimuthal angle $\phi_f$.
$\theta_p$ is the angle between the momenta of incident and scattered photons,
$\vec{k}$ and $\vec{k^\prime}$. The
electron after scattering is not shown in the figure.}
\end{figure}

A review of the calculation of scattered photon energies in the particle-particle
scattering case is in order. 
Figure~\ref{electron_lab_frame} shows the geometry of Compton scattering of an
electron and a photon in a laboratory frame coordinate system $(x_e,y_e,z_e)$ in
which the incident electron with a momentum $\vec{p}$~is moving along the
$z_e$-direction. The incident photon with a momentum $\hbar\vec{k}$
($\hbar$ is the Planck constant) is propagated along the direction with angles 
$(\theta_i, \phi_i)$.
The collision occurs at the origin of the coordinate system. After the
collision, the photon with a momentum $\hbar\vec{k}^\prime$ is scattered into
the direction of $(\theta_f, \phi_f)$.

According to the conservation of the 4-momenta before and after scattering, we
can have 
\begin{equation}
p+ k=p^\prime+ k^\prime,
\label{conservation_law}
\end{equation}
where $p=(E_e/c, \vec{p})$ and $k=(E_p/c,\hbar \vec{k})$ are the 4-momenta of
the electron and photon before the scattering, respectively;
$p^\prime=(E^\prime_e/c, \vec{p}^\prime)$ and $k^\prime=(E_g/c,\hbar
\vec{k}^\prime)$ are their 4-momenta after the scattering; $E_e$ and $E_p$ are
the energies of the electron and photon before the scattering; $E_e^\prime$ and
$E_g$ are their energies after the scattering; and $c$ is the speed of light.
Squaring both sides of Eq.~(\ref{conservation_law}) and following some simple
manipulations, we can obtain the scattered photon energy as follows,
\begin{equation}
E_g = \frac{(1-\beta \cos\theta_i)E_p}{(1-\beta
\cos\theta_f)+(1-\cos\theta_p)E_p/E_e},
\label{scatteredphotonenergy}
\end{equation}
where 
$\beta = v/c$ is the speed of the incident electron relative to the speed of
light, and $\theta_p$ is the angle between the momenta of the incident and
scattered
photons (Fig.~\ref{electron_lab_frame}).

\begin{figure}
\centering
\includegraphics[width=\columnwidth]{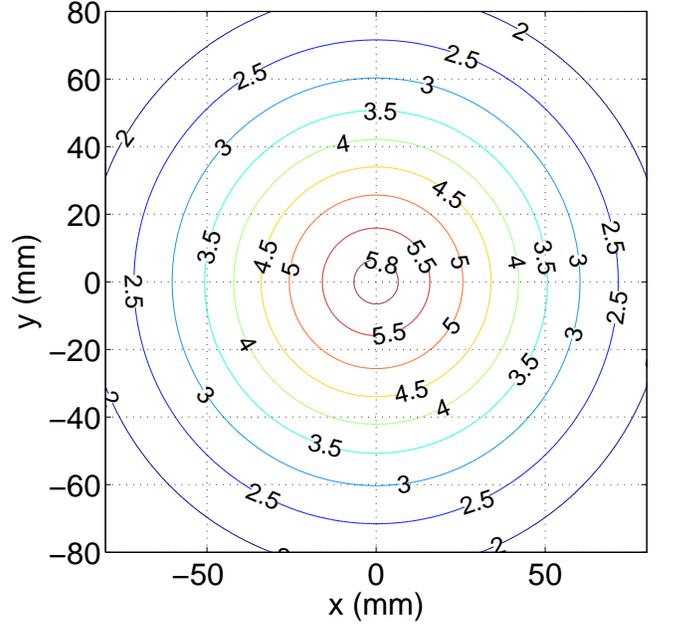}
\caption{\label{sim_en_sp_dist}The relation between the scattered photon energy
(in MeV)
and scattering angle in an observation plane, which is $60$~meters downstream
from the collision point. The scattered photons are produced by $800$~nm photons
scattering with $500$~MeV electrons. Each concentric circle is an equi-energy 
contour curve of the energy distribution of scattered photons.}
\end{figure}

For a head-on collision, $\theta_i = \pi$ and $\theta_p = \pi-\theta_f$,
Eq.~(\ref{scatteredphotonenergy}) can be simplified to 
\begin{equation}
 E_g = \frac{(1+\beta)E_p}{(1-\beta\cos\theta_f)+(1+\cos\theta_f)E_p/E_e}.
\label{energy_head_on}
\end{equation}
Clearly, given the energies of the incident electron and photon, $E_e$ and
$E_p$, the scattered photon energy $E_g$ only depends on the scattering angle
$\theta_f$, independent of the azimuth angle $\phi_f$.
The relation between the scattered photon energy $E_g$ and scattering angle
$\theta_f$ is demonstrated in Fig.~\ref{sim_en_sp_dist}. In this figure, the
scattered photon energies $E_g$ are indicated by the quantities associated with
the concentric circles in the observation plane, and the scattering angles
$\theta_f$ are represented by the radii $R$ of the circles, i.e, $\theta_f =
R/L$, where $L=60$~meters is the distance between the collision point and the
observation plane.  
We can see that the scattered photons with higher energies are concentrated
around
the center ($\theta_f = 0$), while lower energy photons are distributed away
from
the center.
Such a relation, in principle, allows the formation of a scattered photon beam
with a small energy-spread using a simple geometrical collimation technique. 

For a small scattering angle ($\theta_f\ll1$) and an ultra-relativistic electron
($\gamma\gg1$), Eq.~($\ref{energy_head_on}$) can
be simplified to 
\begin{equation}
E_g \approx \frac{4\gamma^2E_p}{1+\gamma^2\theta_f^2+4\gamma^2 E_p/E_e},
\label{scatteredphotonenergy_headon}
\end{equation}
where $\gamma = E_e/(mc^2)$ is the Lorentz factor of the electron and $mc^2$ is
its rest energy.
When the photon is scattered into the backward direction of the incident photon
(i.e., $\theta_f = 0$, sometimes called backscattering), the scattered photon
energy will reach the maximum value given by 
\begin{equation}
E_g^{max} = \frac{4\gamma^2E_p}{1+4\gamma^2 E_p/E_e}.
\label{max_energy_compton}
\end{equation}
Neglecting the recoil effect, i.e., $4\gamma^2 E_p/E_e\ll1$,
Eq.~(\ref{max_energy_compton}) can be reduced to the result given by the
\textit{relativistic Thomson scattering} theory~\cite{Brown:2004zz}
\begin{equation}
 E_g^{max} \approx 4\gamma^2E_p.
\label{energy_thomson}
\end{equation}
We can see that the incident photon energy $E_p$ is boosted by a factor of
approximately $4\gamma^2$ after the backscattering. Therefore, the Compton
scattering of photons with relativistic electrons can be used to produce 
high energy photons, i.e., gamma-ray photons.

Under a set of conditions $\theta_i \approx \pi$ and $\theta_f\approx 0$, the
uncertainties of the scattered photon energy $E_g$ due to the uncertainties of
the variables ($E_e,~E_p,~\theta_f$ and $\theta_i$) in
Eq.~(\ref{scatteredphotonenergy}) can be estimated~\cite{Park_thesis,Park_paper}. 
For example, the relative
uncertainty of the scattered photon energy $\Delta E_g/E_g$ due to the
uncertainty of the electron beam energy $\Delta E_e/E_e$ is given by taking the
derivative of Eq.~(\ref{scatteredphotonenergy}) with respect to $E_e$, i.e.,
\begin{equation}
\frac{\Delta E_g}{ E_g} \approx
2(1-\frac{2\gamma^2E_p/E_e}{1+4\gamma^2E_p/E_e})\frac{\Delta E_e}{E_e}\approx 2
\frac{\Delta E_e}{E_e}.
\end{equation}
Contributions to $\Delta E_g/E_g$ associated with other variables are summarized
in Table~\ref{depedences}.

\begin{table}
\centering
\caption{\label{depedences}Relative uncertainty of the scattered photon energy
$\Delta E_g/ E_g$ due to the uncertainties of various variables in
Eq.~(\ref{scatteredphotonenergy}) under assumptions of $\theta_i \approx \pi$
and $\theta_f\approx 0$.}
\begin{tabular}{c c c c}
\hline\hline
Variables  &  Contributions & Approximated contributions\\
\hline
 $E_e$  & $2(1-\frac{2\gamma^2E_p/E_e}{1+4\gamma^2E_p/E_e})\frac{\Delta
E_e}{E_e} $ & 2$\frac{\Delta E_e}{E_e}$ \\
 $E_p$  &$\frac{1}{1+4\gamma^2E_p/E_e}\frac{\Delta E_p}{E_p}$ & $\frac{\Delta
E_p}{E_p}$ \\
$\theta_f$  & $-\frac{\gamma^2}{1+4\gamma^2E_p/E_e}\Delta \theta_f^2$ &
$-\gamma^2\Delta \theta_f^2$ \\
  $\theta_i$ &$-\frac{\beta}{4}\Delta \theta_i^2$& $-\frac{1}{4}\Delta
\theta_i^2$\\
\hline\hline
\end{tabular}
 \end{table}

\subsection{Scattering cross section}

\subsubsection{Lorentz invariant form}
The general problem concerning the collision is to find the probabilities of
final states for a given initial state of the system, i.e., the scattering cross
section. 
Using Quantum Electrodynamics (QED) theory, the Compton scattering cross section
in the Lorentz invariant form has been calculated
in~\cite{QED_landau,Grozin_book,Grozin_paper}, and the result for unpolarized
electrons scattering with polarized photons is given by
\begin{widetext}
\begin{eqnarray}
\frac{\mathrm{d}\sigma}{\mathrm{d}Y\mathrm{d}\phi_f}&=&\frac{2
r^2_e}{X^2}\left\lbrace
\left(\frac{1}{X}-\frac{1}{Y}\right)^2+\frac{1}{X}-\frac{1}{Y}+\frac{1}{4}
\left(\frac{X}{Y}+\frac{Y}{X}\right)-(\xi_3+\xi^\prime_3)\left[\left(\frac{1}{X}
-\frac{1}{Y}\right)^2+\frac{1}{X}-\frac{1}{Y}\right]\right.\nonumber\\
&&\!\!\!\!\!\!\!\!\!\left.+\xi_1\xi^\prime_1\left(\frac{1}{X}-\frac{1}{Y}+\frac{
1}{2}\right)+\xi_2\xi^\prime_2\frac{1}{4}\left(\frac{X}{Y}+\frac{Y}{X}
\right)\left(1+\frac{2}{X}-\frac{2}{Y}\right)+\xi_3\xi^\prime_3\left[\left(\frac
{1}{X}-\frac{1}{Y}\right)^2+\frac{1}{X}-\frac{1}{Y}+\frac{1}{2}\right]
\right\rbrace,
\label{covariant_crosssection}
\end{eqnarray}
\end{widetext}
where $r_e$ is the \textit{classical electron radius};
$\phi_f$ is the azimuthal angle of the scattered photon; 
$\xi_{1,2,3}$ and $\xi^\prime_{1,2,3}$ are Stokes parameters describing the
incident and scattered photon polarizations in their respective coordinate
systems; $X$ and $Y$ are the Lorentz
invariant variables defined as follows
\begin{equation}
X = \frac{s-(mc)^2}{(mc)^2},~Y = \frac{(mc)^2-u}{(mc)^2},
\label{invariant_quantities}
\end{equation}
where $s$ and $u$ are the \textit{Mandelstam variables}~\cite{QED_landau} given
by
\begin{equation}
s=(p+k)^2,~u=(p-k^\prime)^2.
\end{equation}
$X$ and $Y$ satisfy the inequalities~\cite{QED_landau} 
\begin{equation}
\frac{X}{X+1} \leq Y \leq X.
\label{inequality}
\end{equation}

Since the scattering cross section of Eq.~(\ref{covariant_crosssection}) is
expressed in the Lorentz invariants, it can easily be expressed in terms of the
collision parameters defined in any specific frame of reference.

\subsubsection{Polarization description in lab frame}

\begin{figure}
\centering
\includegraphics[width=\columnwidth]{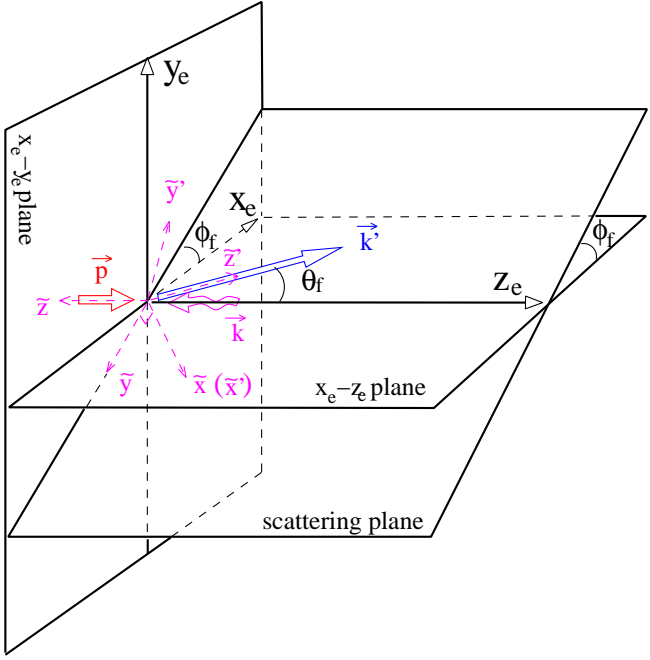}
\caption{\label{scattering plane}Coordinate systems of Compton scattering of an
electron and a photon in a laboratory frame. ($x_e,y_e,z_e$) is the coordinate
system for the incident electron ($\vec{p}$) moving along the $z_e$-axis direction.
For the head-on collision, the incident photon ($\vec{k}$) comes along the negative
$z_e$-axis, and the scattered photon ($\vec{k}^\prime$) is moving along the direction given by the polar angle
$\theta_f$ and azimuthal angle $\phi_f$. The momentum vectors $\vec{k}$ and $\vec{k}^\prime$ 
form the scattering plane. $(\tilde{x}, \tilde{y}, \tilde{z})$ is a right-hand
coordinate system attached to the scattering plane. The $\tilde{z}$-axis is
along the direction of $\vec{k}$; $\tilde{x}$-axis is perpendicular to the
scatter plane; and
$\tilde{y}$-axis is in the scattering plane.
$(\tilde{x}^\prime, \tilde{y}^\prime, \tilde{z}^\prime)$ is another right-hand
coordinate system attached to the scattering plane.
The $\tilde{z}^\prime$-axis is along the direction of $\vec{k}^\prime$;
$\tilde{x}^\prime$-axis is the same as the $\tilde{x}$-axis; 
and $\tilde{y}^\prime$-axis lies in the scattering plane.}
\end{figure}

In the laboratory frame, three right-hand coordinate systems are used in 
Eq.~(\ref{covariant_crosssection}) to 
describe the motion and polarization of the incident electron $(x_e,y_e,z_e)$, the incident 
photon $(\tilde{x}, \tilde{y}, \tilde{z})$, and the scattered photon 
$(\tilde{x}^\prime, \tilde{y}^\prime,\tilde{z}^\prime)$ (Fig.~\ref{scattering plane}).  
The coordinate system $(x_e,y_e,z_e)$ is fixed in the lab frame,
and its $z_e$-axis is along the incident direction of the electron. 
$(\tilde{x}, \tilde{y}, \tilde{z})$ and $(\tilde{x}^\prime, \tilde{y}^\prime,
\tilde{z}^\prime)$ are the local coordinate systems attached to the scattering plane formed 
by the momenta of the incident and scattered photons, $\vec{k}$ and $\vec{k}^\prime$.  
For $(\tilde{x}, \tilde{y}, \tilde{z})$, the $\tilde{x}$-axis is perpendicular to the scattering
plane; the $\tilde{y}$- and $\tilde{z}$-axes are in the scattering plane with 
the $\tilde{z}$-axis along the direction of $\vec{k}$. For 
$(\tilde{x}^\prime,\tilde{y}^\prime,\tilde{z}^\prime)$, 
the $\tilde{x}^\prime$-axis is the same as the $\tilde{x}$-axis for the incident photon,
perpendicular to the scattering plane; and 
the $\tilde{z}^\prime$-axis is along the direction of $\vec{k}^\prime$. 
                   
The Stokes parameters $\xi^{(\prime)}_{1,2,3}$ of the incident and scattered photons in 
Eq.~(\ref{covariant_crosssection}) are defined in their local coordinate systems, respectively.
The parameter $\xi^{(\prime)}_3$ describes the
linear polarization of the photon along the $\tilde{x}^{(\prime)}$- or
$\tilde{y}^{(\prime)}$-axis; the parameter
$\xi_1^{(\prime)}$ describes the linear polarization along the direction at
$\pm45^\circ$ angles relative to the $\tilde{x}^{(\prime)}$-axis; and the parameter 
$\xi_2^{(\prime)}$ represents the degree of circular
polarization of the photon. 

The polarization of the photon is always defined in its local coordinate
system with its momentum being one of the axes. For Compton scattering described by 
Eq.~(\ref{covariant_crosssection}), these 
local coordinate systems $(\tilde{x}, \tilde{y}, \tilde{z})$ and $(\tilde{x}^\prime, \tilde{y}^\prime,
\tilde{z}^\prime)$ are different for different scattering planes. 
However, for the cases that the photons and electrons collide nearly head on to
produce high-energy photons with small scattering angles,
it becomes possible to conveniently express in an approximate manner the
polarization of the incident and scattered photons using a fixed coordinate
system, for example, the lab-frame electron coordinate system $(x_e, y_e, z_e)$.

Let us consider the incident photon with its $\tilde{z}$-axis approximately
parallel to the negative $z_e$-axis.
The Stokes parameter of the incident photon can be related to the
degrees of polarization defined in the fixed electron coordinate system through
the following equations~\cite{CFT_landau,ginzburg},
\begin{eqnarray}
\xi_1 &\approx&  P_t \sin (2\tau-2\phi_f)), \nonumber\\
\xi_2 &\approx&  P_c, \nonumber\\
\xi_3 &\approx& -P_t \cos (2\tau-2\phi_f)),
\label{stokes_lab_initial_photon}
\end{eqnarray}
where $P_t$ and $P_c$ are the degree of linear and circular polarizations of the 
incident photon defined in the coordinate system $(x_e,y_e,z_e)$, respectively; 
$\tau$ is the azimuthal angle of the linear polarization $P_t$ with respect to the $x_e$-axis;
and $\phi_f$ is the azimuthal angle of the scattering plane.

For Compton scattering involving an ultra-relativistic electron, scattered
photons are concentrated in a small scattering angle
($\theta_f < 1/\gamma$).
For these high-energy photons with small scattering angles,
their $\tilde{z}^\prime$-axes are approximately parallel to the $z_e$-axis.
Neglecting the polar angle (i.e. $\theta_f \ll 1$), the Stokes parameters of the
scattered photon can be expressed approximately
using a set of Stokes parameters defined in the fixed electron
coordinate system as~\cite{ginzburg},
\begin{eqnarray}
\xi^\prime_1&\approx&-\bar{\xi}^\prime_1\cos2\phi_f+\bar{\xi}^\prime_3\sin2\phi_f,
\nonumber\\
~\xi^\prime_2&\approx&\bar{\xi}^\prime_2,\nonumber\\
~\xi^\prime_3&\approx&-\bar{\xi}^\prime_1\sin2\phi_f-\bar{\xi}^\prime_3\cos2\phi_f,
\label{final_photon_stokes}
\end{eqnarray}
where $\bar{\xi}'_{1,2,3}$ are the Stokes parameters defined in
the coordinate system $(x_e, y_e, z_e)$.

\subsection{Spatial and energy distributions of scattered photons}
Based upon Eqs.~(\ref{covariant_crosssection}),
(\ref{stokes_lab_initial_photon})~and~(\ref{final_photon_stokes}), we can
calculate the spatial and energy distributions of a gamma-ray beam produced by
Compton scattering of a monoenergetic electron and laser beams with zero
transverse beam sizes, i.e., the particle-particle scattering. 

Let us consider Compton scattering of an unpolarized electron and a polarized
laser photon without regard to their polarizations after the scattering. The
differential cross section is obtained by setting $\xi^\prime_{1,2,3}$ to zero
in Eq.~(\ref{covariant_crosssection}) and multiplying the result by a factor of two
for the summation over the polarizations of the scattered
photons~\cite{QED_landau}. Thus, the differential cross section is given
by~\cite{Park_paper}
\begin{eqnarray}
\frac{\mathrm{d}\sigma}{\mathrm{d}Y\mathrm{d}\phi_f} & = &
\frac{4r^2_e}{X^2}\left\lbrace 
(1-\xi_3)\left[\left(\frac{1}{X}-\frac{1}{Y}\right)^2+\frac{1}{X}-\frac{1}{Y}
\right]\right.\nonumber\\
&&\left. +\frac{1}{4}\left(\frac{X}{Y}+\frac{Y}{X}\right)\right\rbrace .
\label{angular_dif_crosssection}
\end{eqnarray}

The total cross section can be obtained by integrating
Eq.~(\ref{angular_dif_crosssection}) with respect to $Y$ and $\phi_f$, 
\begin{eqnarray}
 \sigma_{tot} &=& 2\pi
r_e^2\frac{1}{X}\left\lbrace\left(1-\frac{4}{X}-\frac{8}{X^2}
\right)\log(1+X)\right.\nonumber\\
&&\left.+\frac{1}{2} +\frac{8}{X}-\frac{1}{2(1+X)^2}\right\rbrace.
\label{tot_scat_cross}
\end{eqnarray}
Note that the Stokes parameter $\xi_3$ depends on $\phi_f$; however, after
integration over $\phi_f$ the dependence vanishes.

Neglecting the recoil effect ($X\ll1$),  we can have
\begin{equation}
  \sigma_{tot} = \frac{8\pi r_e^2}{3}(1-X)\approx  \frac{8\pi r_e^2}{3},
\label{total_cross_section}
\end{equation}
which is just the \textit{classical Thomson cross section}.

\subsubsection{Spatial distribution}
For a head-on collision ($\theta_i=\pi$) in a laboratory frame, according to
Eq.~(\ref{invariant_quantities}) the Lorentz invariant quantities $X$ and $Y$
are given by  
\begin{equation}
 X = \frac{2\gamma E_p(1+\beta)}{mc^2},~Y=\frac{2\gamma
E_g(1-\beta\cos\theta_f)}{mc^2},
\label{invariant_lab_frame}
\end{equation}
and 
\begin{equation}
\mathrm{d}Y=2 \left(\frac{E_g}{mc^2}\right)^2\sin\theta_f\mathrm{d}\theta_f.
\label{dy_domega}
\end{equation}
Substituting $\mathrm{d}Y$ into Eq.~(\ref{angular_dif_crosssection}), the
angular
differential cross section is given by
\begin{eqnarray}
\frac{\mathrm{d}\sigma}{\mathrm{d}\Omega}\!\!\! &=&
\!\!\!\frac{8r^2_e}{X^2}\!\!\left\lbrace 
\![1\!+\!P_t\cos(2\tau\!-\!\!2\phi_f)]\!\!\left[\left(\frac{1}{X}-\frac{1}{Y}
\right)^2\!\!+\!\!\frac{1}{X}\!-\!\frac{1}{Y}\right]\right.\nonumber\\
&&\left.+\frac{1}{4}\left(\frac{X}{Y}+\frac{Y}{X}\right)\right\rbrace
\left(\frac{E_g}{mc^2}\right)^2.
\label{crosssection-1}
\end{eqnarray}
where $\mathrm{d}\Omega = \sin\theta_f \mathrm{d}\theta_f \mathrm{d}\phi_f$ and
$\xi_3$ has been expressed in terms of $P_t$
(Eq.~(\ref{stokes_lab_initial_photon})). 

\begin{figure*}
\centering
\begin{tabular}{cc}
\includegraphics[width=3in]{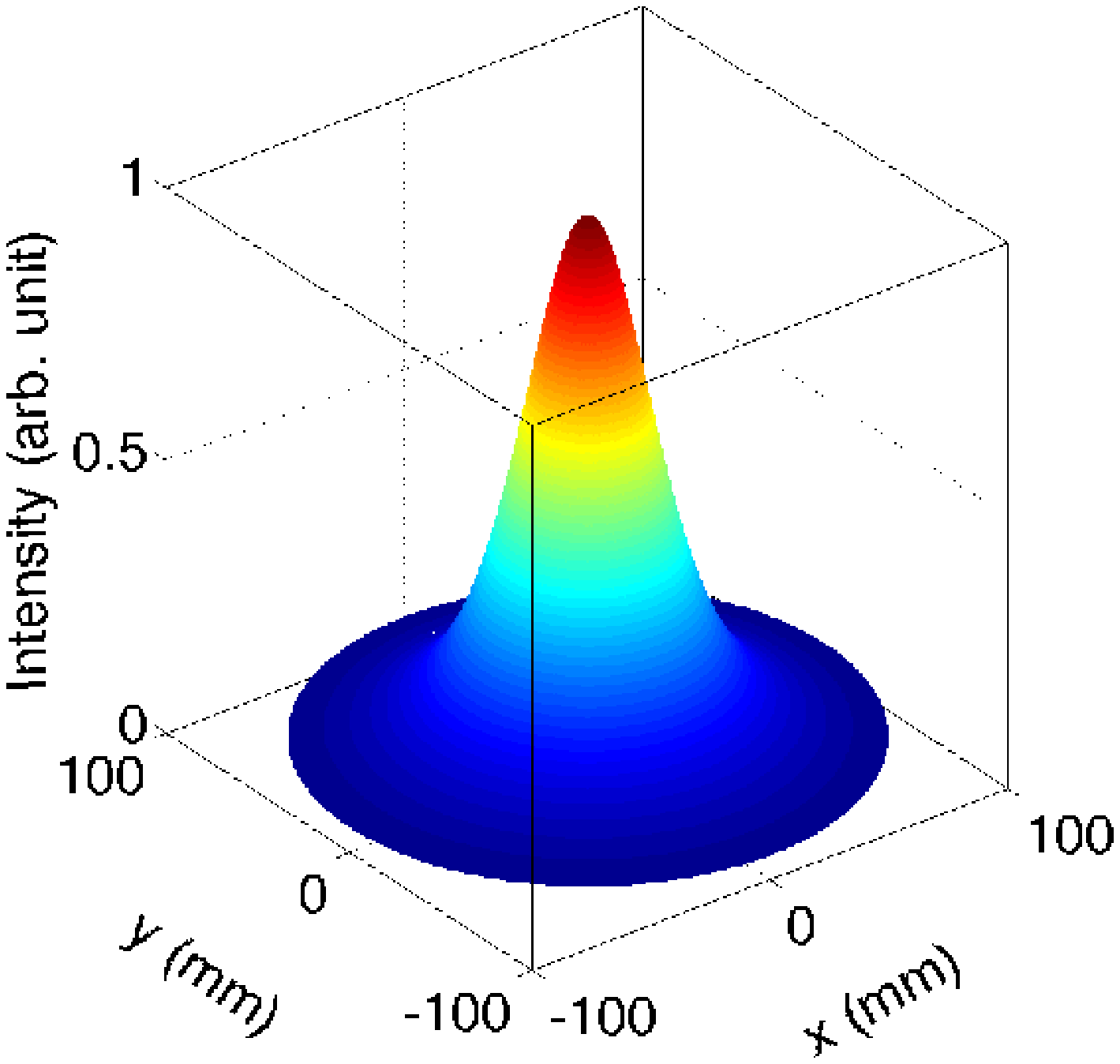}&
\includegraphics[width=3in]{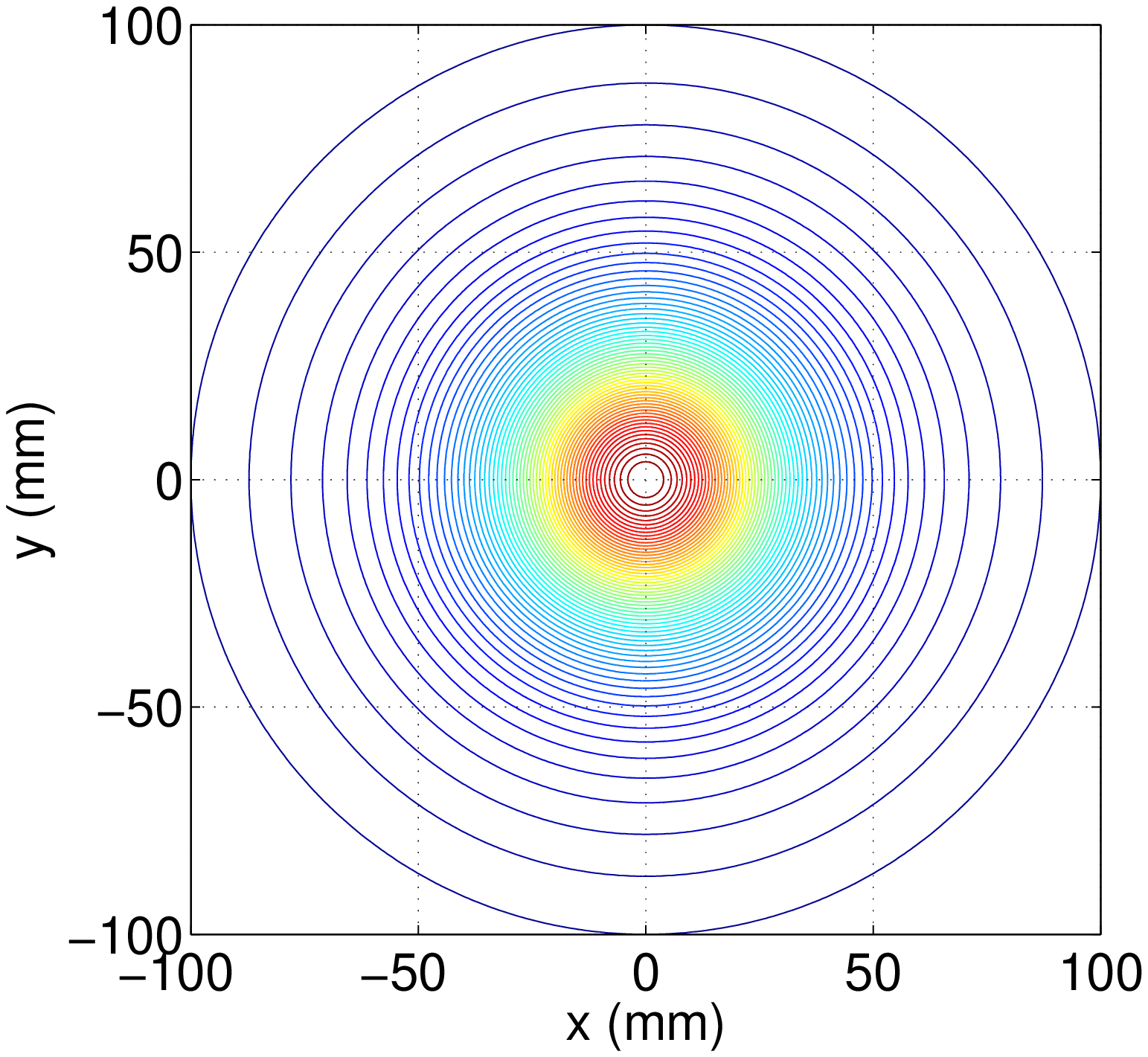} 
\end{tabular}
\caption{\label{cir_dist}(Color) The computed spatial distribution of Compton gamma-ray
photons produced by a head-on collision of a circularly polarized $800$~nm laser
beam with an unpolarized $500$~MeV electron beam. The distribution is calculated
for a location $60$~meters downstream from the collision point. The left plot is
a 3-dimensional intensity distribution, and the right plot is the contour plot
of the gamma-beam intensity distribution.}
\end{figure*}

\begin{figure*}
\centering
\begin{tabular}{cc}
\includegraphics[width=3in]{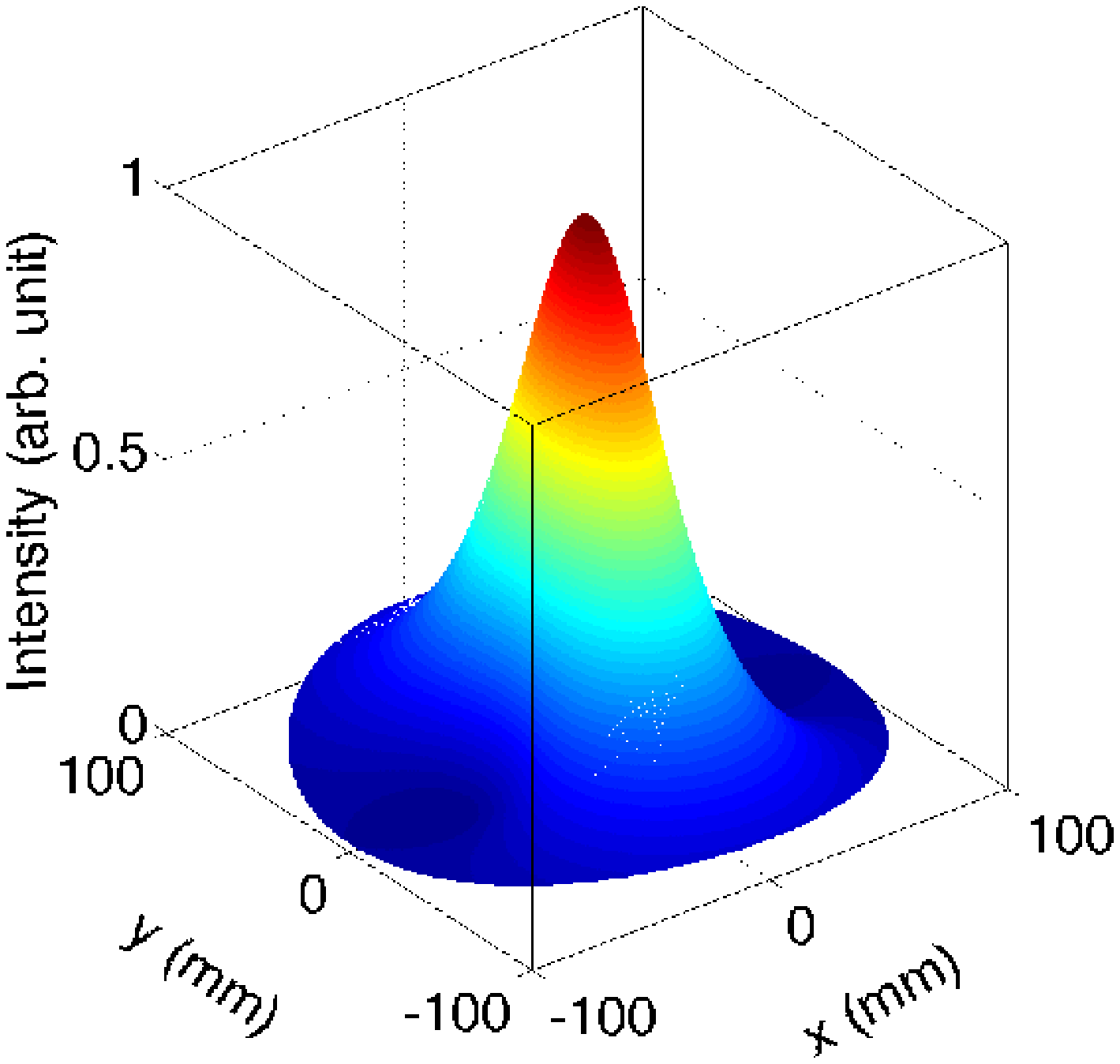} &
\includegraphics[width=3in]{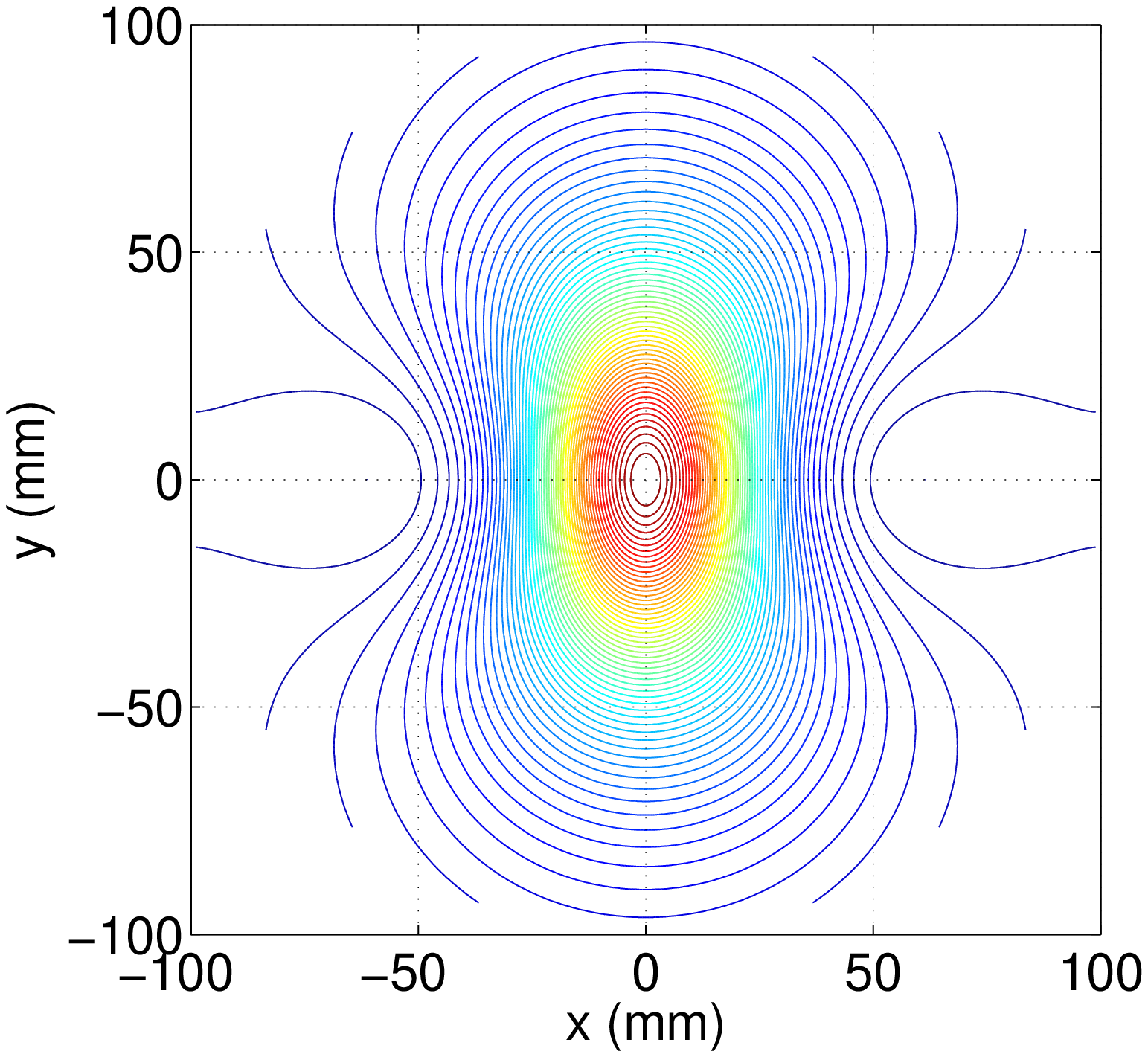}
\end{tabular}
\caption{\label{linear_dist}(Color) The computed spatial distribution of Compton
gamma-ray photons produced by a head-on collision of a linearly polarized
$800$~nm laser beam with an unpolarized $500$~MeV electron beam. The
polarization of the incident photon beam is along the horizontal direction. The
distribution is calculated for a location $60$~meters downstream from the
collision point. The left plot is a 3-dimensional intensity distribution, and
the right plot is the contour plot of the gamma-beam intensity distribution.}
\end{figure*}

From Eq.~(\ref{crosssection-1}), we can see that the differential cross section
depends on the azimuthal angle $\phi_f$ of the scattered photon through the term
$P_t\cos(2\tau-2\phi_f)$.  
For a circularly polarized or unpolarized incident photon beam ($P_t = 0$), this
dependency vanishes. Therefore, the distribution of scattered photons is
azimuthally symmetric. 
However, for a linearly polarized incident photon beam ($P_t \neq 0$), the
differential cross section is azimuthally modulated, and the gamma photon
distribution is azimuthally asymmetric.   
Figs.~\ref{cir_dist} and~\ref{linear_dist} illustrate the spatial distributions
of Compton gamma photons at a location $60$~meters downstream from the collision
point for both circularly and linearly polarized incident photon beams. 
In these figures we can also see that the distribution of scattered photons peaks
sharply 
along the direction of the incident electron beam. This
demonstrates that the gamma-ray photons produced by Compton scattering of a
relativistic electron beam and a laser beam are mostly scattered into the
electron beam direction within a narrow cone.

\subsubsection{Energy distribution}
For a head-on collision in the laboratory frame, it can be shown that
\begin{equation}
 Y = X\frac{\beta E_e-E_g}{\beta E_e-E_p}.
\end{equation}
Thus, 
\begin{equation}
 \mathrm{d}Y = -X\frac{\mathrm{d} E_g}{\beta E_e-E_p}. 
\label{dy_dEg}
\end{equation}
Substituting $\mathrm{d}Y$ in Eq.~(\ref{angular_dif_crosssection}) and
integrating the result with respect to the azimuth angle $\phi_f$, we can obtain
the energy distribution of scattered photons as follows
\begin{eqnarray}
\frac{\mathrm{d}\sigma}{\mathrm{d}E_g } & = &\frac{8 \pi r^2_e}{X(\beta
E_e-E_p)} 
\left[\left(\frac{1}{X}-\frac{1}{Y}\right)^2+\frac{1}{X}-\frac{1}{Y}
\right.\nonumber\\
&&\left.+\frac{1}{4}\left(\frac{X}{Y}+\frac{Y}{X}\right)\right].
\label{crosssection-en}
\end{eqnarray}

\begin{figure}
\centering
\includegraphics[width=\columnwidth]{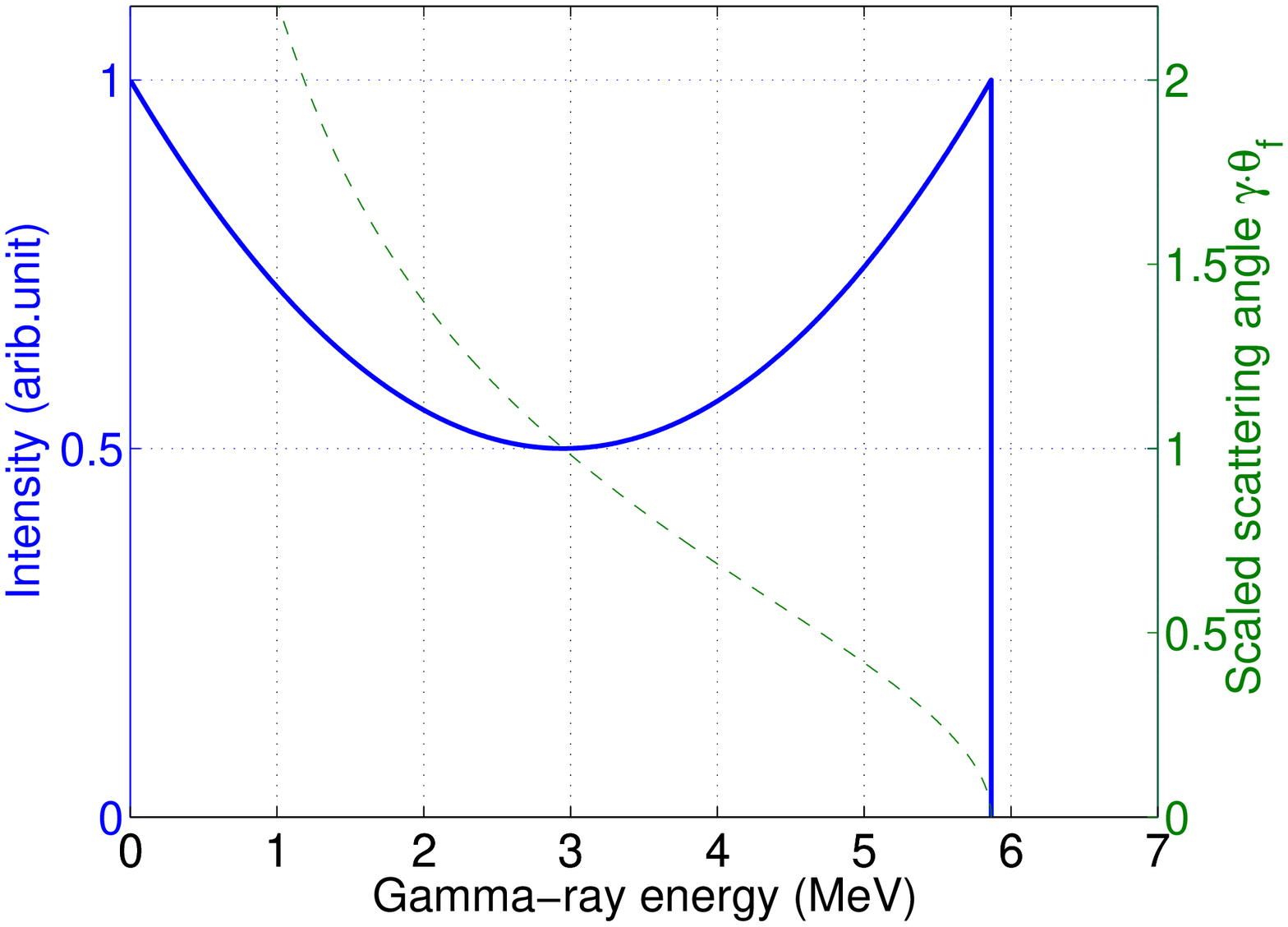}
\caption{\label{sim_en_dist}The computed energy distribution of Compton gamma-ray photons
produced by a head-on collision of a $800$~nm laser beam with a $500$~MeV
electron beam. The scaled scattering angle $\gamma\theta_f$ with the electron
Lorentz factor versus the gamma-ray photon energy is also shown in the plot. The
solid line represents the energy distribution of the gamma-ray photons, and the
dash line represents the relation between the scaled scattering angle and photon
energy.}
\end{figure}

The energy spectrum calculated using Eq.~(\ref{crosssection-en}) is shown in
Fig.~\ref{sim_en_dist}. The spectrum has a high energy cutoff edge which is
determined by the incident electron and photon energies according to
Eq.~(\ref{max_energy_compton}). In Fig.~\ref{sim_en_dist}, we can see the
spectral intensity has a maximum value at the scattering angle $\theta_f=0$, and
a minimum value around the scattering angle $\theta_f = 1/\gamma$. The ratio
between them is about $2$ when the recoil effect is negligible. This will be
shown in the next section.

Note that the energy spectrum shown in Fig.~\ref{sim_en_dist} is for a Compton
gamma-ray beam without collimation. However, if the gamma-ray beam is collimated
by a round aperture with a radius of $R$ and distance $L$ from the collision
point, the energy spectrum will have a low energy cutoff edge, and its value can
be calculated using Eq.~(\ref{scatteredphotonenergy_headon}) with $\theta_f =
R/L$.

\subsubsection{Observations for a small recoil effect}
For a small recoil effect ($X\ll1$), we can approximate
Eqs.~(\ref{crosssection-1}) and~(\ref{crosssection-en}) to draw several useful
conclusions.

For convenience, we first define 
\begin{equation}
 f(Y) =
\left(\frac{1}{X}-\frac{1}{Y}\right)^2+\frac{1}{X}-\frac{1}{Y}+\frac{1}{4}
\left(\frac{X}{Y}+\frac{Y}{X}\right).
\label{f_term}
\end{equation}
Using the inequality Eq.~(\ref{inequality}), it can be found that 
\begin{equation}
\frac{1}{4(1+X)} \leq f(Y) \leq \frac{2+X}{4},
\end{equation}
approximately (with a negligible recoil effect, $X\ll1$),
\begin{equation}
\frac{1}{4}\leq f(Y)\leq \frac{1}{2}. 
\end{equation}

Thus, the maximum and minimum spectral flux of the Compton gamma-ray beam are
given by  
\begin{equation}
 (\frac{\mathrm{d}\sigma}{\mathrm{d}E_g})_{max} = \frac{8 \pi r^2_e}{X(\beta
E_e-E_p)}\frac{2+X}{4},
\label{maximum_spectrum_intensity}
\end{equation}
and 
\begin{equation}
 (\frac{\mathrm{d}\sigma}{\mathrm{d}E_g})_{min} = \frac{8 \pi r^2_e}{X(\beta
E_e-E_p)}\frac{1}{4(1+X)}.
\end{equation}
The ratio between them is  
\begin{equation}
 \frac{(\mathrm{d}\sigma/\mathrm{d}E_g)_{max}}{(\mathrm{d}\sigma/\mathrm{d}E_g)_
{min}} = (2+X)(1+X)\approx 2,
\end{equation}
which is shown in Fig.~\ref{sim_en_dist}.

When $\theta_f = 0$, we can have 
\begin{equation}
 E_g \approx 4\gamma^2E_p,~Y \approx X(1-X).
\end{equation}
Substituting $Y$ in Eq.~(\ref{f_term}), we have $f(Y) \approx 1/2$. Thus, the
spectral flux has a maximum value around the scattering angle $\theta_f = 0$. 
When $\theta_f = 1/\gamma$, we can have 
\begin{equation}
 E_g \approx 2\gamma^2E_p,~Y \approx X(1-\frac{X}{2}).
\end{equation}
Substituting $Y$ into Eq.~(\ref{f_term}), we have $f(Y)\approx 1/4$. Therefore,
the spectral flux has a minimum value around the scattering angle $\theta_f =
1/\gamma$. These results are illustrated in Fig.~\ref{sim_en_dist}.

Expressed in terms of the total scattering cross section of
Eq.~(\ref{total_cross_section}), the fraction of scattered photons in the energy
range $[E_g^{max}-\Delta E_g^{max}, E_g^{max}]$ 
can be found approximately as
\begin{equation}
 \frac{\Delta \sigma_{max}}{\sigma_{tot}} \approx 
\frac{3(2+X)}{4(1-X)}\frac{\Delta E_g^{max}}{E_g^{max}}\approx 1.5\frac{\Delta
E_g^{max}}{E_g^{max}}.
\end{equation}
This is a simple formula which can be used to estimate the portion of the total
gamma-ray flux with a desirable energy spread $\Delta E_g^{max}$ after
collimation. 

For a circularly polarized or unpolarized incident photon beam, according to
Eq.~(\ref{crosssection-1}), it can also be calculated that the
angular intensity of scattered gamma-ray photons at the scattering angle
$\theta_f = 1/\gamma$ is about $1/8$
of the maximum intensity at the scattering angle $\theta_f = 0$,
i.e., 
\begin{equation}
\frac{(\mathrm{d}\sigma/\mathrm{d}\Omega)_{\theta_f=1/\gamma}}{(\mathrm{d}
\sigma/\mathrm{d}\Omega)_{\theta_f=0}}\approx\frac{1}{8}.  
\end{equation}

In addition, integrating Eq.~(\ref{angular_dif_crosssection}) over the entire
solid angle of the cone with a half-opening angle of $1/\gamma$, i.e., integrating $Y$
over the range of $X(1-X/2) \leqslant Y \leqslant X(1-X)$ and $\phi_f$ over the
range from $0$ to $2\pi$, we can have 
\begin{equation}
\sigma_1 =
\int_0^{2\pi}\mathrm{d}\phi\int_0^{1/\gamma}\frac{\mathrm{d}\sigma}{\mathrm{d}
\Omega}\sin\theta \mathrm{d}\theta\approx\frac{4\pi
r_e^2}{3}=\frac{1}{2}\sigma_{tot}.
\label{flux_in_cone}
\end{equation}
Comparing Eq.~(\ref{flux_in_cone}) to the total cross section of
Eq.~(\ref{total_cross_section}), we can conclude that about half of the total
gamma-ray photons are scattered into the $1/\gamma$ cone. This can be explained
by considering the Compton scattering in the electron rest frame. In this frame,
the Compton scattering process is just like ``dipole'' radiation: the gamma-ray
photons are scattered in all the directions, a half of the gamma photons is
scattered into the forward direction, and the other half into the backward
direction. When transformed to the laboratory frame, the gamma-ray photon
scattered into the forward direction in the rest frame will be concentrated in
the $1/\gamma$ cone in the laboratory frame.

\subsection{Polarization of scattered photons\label{polarization_study}}
For polarized photons scattering with unpolarized electrons without regard to
the final electron polarization, the cross section is given by
Eq.~(\ref{covariant_crosssection}).
Substituting $\xi_{1,2,3}$ and $\xi^\prime_{1,2,3}$ using
Eqs.~(\ref{stokes_lab_initial_photon}) and (\ref{final_photon_stokes}), and
assuming the linear polarization of the incident photon beam is along the
$x_e$-axis, i.e., $\tau = 0$, we can get
\begin{equation}
\frac{\mathrm{d}\sigma}{\mathrm{d}Y\mathrm{d}\phi_f}=\frac{2r^2_e}{X^2}
\left(\Phi_0+\displaystyle\sum_{i=1}^3\Phi_i\bar{\xi}^\prime_i\right),
\label{finalphoton-polar}
\end{equation}
where
\begin{eqnarray}
\Phi_0&=&\left(\frac{1}{X}-\frac{1}{Y}\right)^2+\frac{1}{X}-\frac{1}{Y}+\frac{1}
{4}\left(\frac{X}{Y}+\frac{Y}{X}\right)\nonumber\\
&&+\left[\left(\frac{1}{X}-\frac{1}{Y}\right)^2+\frac{1}{X}-\frac{1}{Y}\right]
P_t\cos2\phi_f,\nonumber\\
\Phi_1&=&\frac{1}{2}\left(\frac{1}{X}-\frac{1}{Y}
+1\right)^2P_t\sin4\phi_f\nonumber\\
&&+\left[\left(\frac{1}{X}-\frac{1}{Y}\right)^2+\frac{1}{X}-\frac{1}{Y}\right]
\sin2\phi_f,\nonumber\\
\Phi_2&=&\frac{1}{4}\left(\frac{X}{Y}+\frac{Y}{X}\right)\left(\frac{2}{X}-\frac{
2}{Y}+1\right)P_c,\nonumber\\
\Phi_3&=&-\left(\frac{1}{X}-\frac{1}{Y}+\frac{1}{2}
\right)P_t\sin^22\phi_f\nonumber\\
&&+\left[\left(\frac{1}{X}-\frac{1}{Y}\right)^2+\frac{1}{X}-\frac{1}{Y}+\frac{1}
{2}\right]P_t\cos^22\phi_f\nonumber\\    
&&+\left[\left(\frac{1}{X}-\frac{1}{Y}\right)^2+\frac{1}{X}-\frac{1}{Y}\right]
\cos2\phi_f.
\end{eqnarray}

It should be noted that the Stokes parameters $\bar{\xi}^\prime_{1,2,3}$
describe the polarization of the scattered photon selected by a detector, not
the polarization of the photon itself~\cite{QED_landau}. In order to distinguish
them from the
detected Stokes parameters $\bar{\xi}^\prime_{1,2,3}$, we denote the Stokes
parameters of the scattered photon itself by $\xi^f_{1,2,3}$. According to the
rules presented in section 65 of~\cite{QED_landau}, $\xi^f_{1,2,3}$ are given by
\begin{equation}
\xi^f_i = \frac{\Phi_i}{\Phi_0},~~i=1,2,3.
\end{equation}
Integrating Eq.~(\ref{finalphoton-polar}) over the azimuthal angle $\phi_f$
gives
\begin{equation}
\frac{\mathrm{d}\sigma}{\mathrm{d}Y}=\frac{2r^2_e}{X^2}\left\lbrace
\langle\Phi_0\rangle+\displaystyle\sum_{i=1}^3\langle\Phi_i\rangle\langle\bar{
\xi}^\prime_i\rangle\right\rbrace,
\end{equation}
where
\begin{eqnarray}
\langle\Phi_0\rangle
&=&2\pi\left[\left(\frac{1}{Y}-\frac{1}{Y}\right)^2+\frac{1}{X}-\frac{1}{Y}
+\frac{1}{4}\left(\frac{X}{Y}+\frac{Y}{X}\right)\right],\nonumber\\
\langle\Phi_1\rangle&=&0,\nonumber\\
\langle\Phi_2\rangle&=&\frac{\pi}{2}\left(\frac{X}{Y}+\frac{Y}{X}
\right)\left(\frac{2}{X}-\frac{2}{Y}+1\right)P_c,\nonumber\\
\langle\Phi_3\rangle&=&\pi\left(\frac{1}{X}-\frac{1}{Y}\right)^2P_t.
\end{eqnarray}
Therefore, the averaged Stokes parameters of the scattered photons over the
angle $\phi_f$ are given by
$\langle\xi^f_i\rangle=\langle\Phi_i\rangle/\langle\Phi_0\rangle$, which
depend on the incident photon polarization and variables $X$ and $Y$.

\begin{figure}
\centering
\includegraphics[width=\columnwidth]{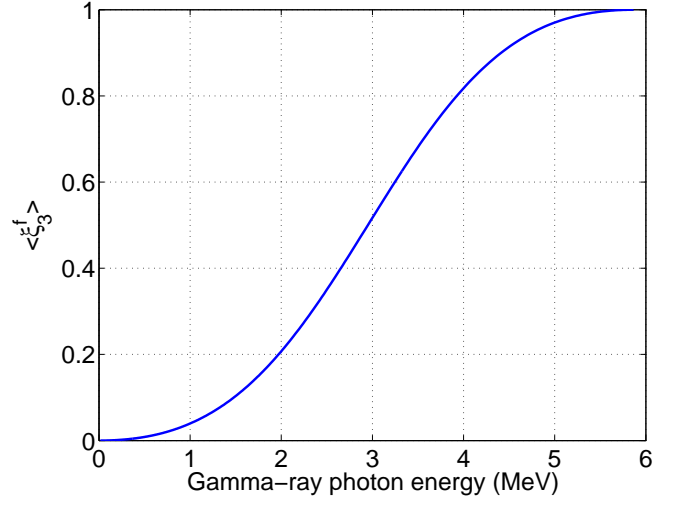}
\caption{The average Stokes parameter $\langle\xi^f_3\rangle$ of Compton
gamma-ray
photons produced by a $100$\% horizontally polarized ($P_t=1, P_c=0,\tau=0$)
$800$~nm 
laser photons head-on colliding with an unpolarized $500$ MeV electrons. }
\label{fig-stokes1}
\end{figure}

For example, for 100\% horizontally polarized ($P_t=1, P_c=0,\tau=0$) incident
photons 
scattering with unpolarized electrons, the
average Stokes parameters of the scattered photons are given by
\begin{eqnarray}
\!\!\!\!\langle\xi^f_1\rangle
&=&\frac{\langle\Phi_1\rangle}{\langle\Phi_0\rangle}=0,
~~~~~~~~~~\langle\xi^f_2\rangle
=\frac{\langle\Phi_2\rangle}{\langle\Phi_0\rangle}=0,\nonumber\\
\!\!\!\!\langle\xi^f_3\rangle
&=&\frac{\langle\Phi_3\rangle}{\langle\Phi_0\rangle}=\frac{2(\frac{1}{X}-\frac{1
}{Y})^2}{4(\frac{1}{X}-\frac{1}{Y})^2+\frac{4}{X}-\frac{4}{Y}+\frac{X}{Y}+\frac{
Y}{X}}.
\end{eqnarray}
Clearly, the scattered photons retain the polarization of the incident photons.
$\langle\xi^f_3\rangle$ as a function of the scattered photon energy is shown in
Fig.~\ref{fig-stokes1} for $800$~nm laser photons head-on colliding with
$500$~MeV electrons.
It can be seen that the average Stokes parameter $\langle\xi^f_3\rangle$ of
scattered gamma-ray photons
is almost equal to 1 around the maximum scattered photon energy as in this case
the recoil effect is negligible. It means the scattered gamma-ray photons with the
maximum energy are almost
$100$\% horizontally polarized.

\section{Beam-beam scattering}  

In the previous section we discussed the spatial and spectral distributions
of a gamma-ray beam produced by Compton scattering of monoenergetic electron
and laser beams with zero transverse beam sizes, i.e., particle-particle
scattering. However, in the reality, the incoming electron and laser beams have
finite spatial and energy distributions, which will change the distributions of
the
scattered gamma-ray beam. Therefore, there remains a need to understand the
characteristics of a Compton gamma-ray beam produced by scattering of a laser
beam and an electron beam with specific spatial and energy distributions, i.e.,
the beam-beam scattering. 

In this section, we discuss the beam-beam Compton scattering process. First, we
derive a
simple formula to calculate the total flux of the Compton gamma-ray beam. Then, we
present two methods, 
a semi-analytical calculation and a Monte Carlo simulation, to study the spatial
and spectral distributions of the gamma-ray beam. 
Based upon these methods, two
computing codes, a numerical integration code and a Monte Carlo simulation code
have
been developed. These two codes have been benchmarked against the experimental
results at High Intensity Gamma-ray Source (HI$\gamma$S) facility at Duke
University. 

\subsection{Geometry of beam-beam scattering}

\begin{figure}
\centering
\includegraphics[width= \columnwidth ]{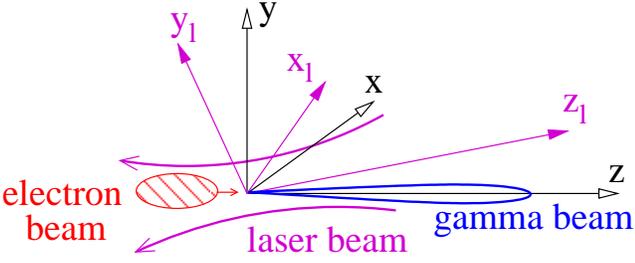}
\caption{\label{lab_laser_frame} Compton scattering of a pulsed electron beam
and a pulsed laser beam in the laboratory frame. Two coordinate systems are
defined to describe electron and laser beams: the first coordinate system
$(x,y,z)$ is the electron-beam coordinate system in which the electron beam is
moving along the $z$-axis direction; the $(x_l,y_l,z_l)$ system is the
laser-beam coordinate system in which the laser beam propagates in the
negative $z_l$-axis direction. The coordinate systems $(x,y,z)$ and
$(x_l,y_l,z_l)$ share the same origin.} 
\end{figure}

Figure~\ref{lab_laser_frame} shows Compton scattering of a pulsed electron beam
and a pulsed laser beam in a laboratory frame. Two coordinate systems are used: 
$(x,y,z)$ for the electron-beam moving along the $z$-direction;
the $(x_l,y_l,z_l)$ for the laser-beam propagating in the
negative $z_l$-direction. These two coordinate systems share a common
origin. The time $t=0$ is chosen for the instant when the centers of the
electron beam and laser pulse arrive at the origin. The definition of these two
coordinate systems allows the study of the Compton scattering process with an
arbitrary collision angle, i.e, the angle between $z$-axis and negative
$z_l$-axis. For a head-on collision, the collision angle equals $\pi$. In this
case, the electron and laser coordinate systems coincide.

In these coordinate systems, the electron and laser beams with Gaussian
distributions in their phase spaces can be described by their respective
intensity functions as follows~\cite{Vladimir}
\begin{widetext}
\begin{eqnarray}
f_e(x,y,z,x^\prime,y^\prime,p,t)\!\!\!&=&\!\!\!\frac{1}{
(2\pi)^3\varepsilon_x\varepsilon_y\sigma_p\sigma_z}\!\exp\!\left[-\frac{
\gamma_xx^2+2\alpha_xxx^\prime+\beta_xx^{\prime2}}{2\varepsilon_x}\!-\!\frac{
\gamma_yy^2+2\alpha_yyy^\prime+\beta_yy^{\prime2}}{2\varepsilon_y}\!-\!\frac{
(p-p_0)^2}{2\sigma^2_p}\!-\!\frac{(z-ct)^2}{2\sigma^2_z}\right],\nonumber\\
f_p(x_l,y_l,z_l,k,t)\!\!\!&=&\!\!\!\frac{1}{4\pi^2\sigma_l\sigma_k\sigma_w^2}
\!\exp\!\left[-\frac{x_l^2+y_l^2}{2\sigma_w^2}-\frac{(z_l+ct)^2}{2\sigma_l^2}
-\frac{(k-k_0)^2}{2\sigma^2_k}\right],~\sigma_w =\sqrt{\frac{\lambda
\beta_0}{4\pi}\left(1+\frac{z_l^2}{\beta_0^2}\right)},
\label{electron-photon-dist}
\end{eqnarray}
\end{widetext}
$p$ is the momentum of an electron, and $p_0$ is the centroid momentum of the
electron beam; $x^\prime$ and $y^\prime$ are the angular divergences of the
electron beam in the $x$- and $y$- directions, respectively;  
$\alpha_{x,y},\beta_{x,y}$ and $\gamma_{x,y}$ are Twiss parameters of the
electron beam; $\sigma_p$, $\sigma_z$ and $\varepsilon_{x,y}$ are the electron
beam momentum spread, RMS bunch length, and transverse emittance, respectively; $k$ and
$\lambda$ are the wavenumber and wavelength of a laser photon, and $k_0$ is the
centroid wavenumber of the laser beam; $\beta_0, \sigma_k$ and $\sigma_l$ are
the Rayleigh range, the RMS energy spread and bunch length of the laser beam.
Note that
the waist of the laser beam is assumed to be at the origin of both coordinate systems.

\subsection{Total flux}
The number of collisions occurring during a time $\mathrm{d}t$ and inside a
phase space volume $\mathrm{d}^3p~\mathrm{d}^3k~\mathrm{d}V$ is given by~\cite{CFT_landau}   
\begin{eqnarray}
\mathrm{d}N(\vec{r},\vec{p},\vec{k},t)&=&\sigma_{tot}(\vec{p},\vec{k})c(1-\vec{
\beta}\cdot\vec{k}/|\vec{k}|)n_e(\vec{r},\vec{p},t)\nonumber\\
&&\times n_p(\vec{r},\vec{k},t)\mathrm{d}^3p~\mathrm{d}^3k~\mathrm{d}V\mathrm{d}t,
\label{collisions_rate}
\end{eqnarray}
where $\sigma_{tot}(\vec{p},\vec{k})$ is the total Compton scattering cross
section which is determined by the momenta of the incident electron and laser
photon, $\vec{p}$ and $\hbar\vec{k}$; $\vec{\beta}=\vec{v}_e/c$ is the relative
velocity of the incident electron;
$n_e(\vec{r},\vec{p},t)= N_e f_e(\vec{r},\vec{p},t)$  and
$n_p(\vec{r},\vec{k},t) = N_p f_p(\vec{r},\vec{k},t)$, where
$f_e(\vec{r},\vec{p},t)$ and $f_p(\vec{r},\vec{k},t)$ are the phase space
intensity functions of electron beam and laser pulse, and $N_e$ and $N_p$ are
the
total numbers of electrons and laser photons in their respective pulses.

To calculate the total number of scattered gamma-ray photons produced by 
collision, Eq.~(\ref{collisions_rate}) needs to be integrated for the entire
phase space and the collision time, i.e.,
\begin{eqnarray}
N_{tot} &=& \int \mathrm{d} N(\vec{r},\vec{p},\vec{k},t)\nonumber\\
&=& N_e N_p\int \sigma_{tot}(\vec{p},\vec{k})c(1-\beta\cos\theta_i)\nonumber\\
&&\times
f_e(\vec{r},\vec{p},t)f_p(\vec{r},\vec{k},t)\mathrm{d}^3p~\mathrm{d}^3k~\mathrm{d}
V\mathrm{d}t.
\label{tot_flux_integration}
\end{eqnarray}
where $\theta_i$ is the collision angle between the incident electron and laser
photon.
Assuming collisions occur at the waists of both beams $(\alpha_x =\alpha_y = 0,
\sigma_w =\sqrt{\lambda \beta_0/(4\pi)} )$, the spatial and momentum phase space
in the density functions 
can be separated, i.e, $f_e(\vec{r},\vec{p},t) =
f_e(\vec{r},t)f_e(\vec{p})$ and $f_p(\vec{r},\vec{k},t) =
f_p(\vec{r},t)f_p(\vec{k})$. Since the cross section
$\sigma_{tot}(\vec{p},\vec{k})$ only depends on $\vec{p}$ and $\vec{k}$,  we can
have
\begin{equation}
N_{tot} = N_e N_p\int  \mathcal{L}_{sc} \sigma_{tot}(\vec{p},\vec{k})
f_e(\vec{p})f_p(\vec{k})\mathrm{d}^3p~\mathrm{d}^3k,
\label{tot_flux_lumin}
\end{equation}
where
\begin{equation}
\mathcal{L}_{sc} = c (1-\beta\cos\theta_i) \int
f_e(\vec{r},t)f_p(\vec{r},t)\mathrm{d}V\mathrm{d}t
\label{luminosity_single}
\end{equation}
is the single-collision luminosity defined as the number of scattering events
produced per unit scattering cross section, which has dimensions of
1/area~\cite{Chao:1999qt}.
For a head-on collision ($\theta_i = \pi$) of a relativistic electron ($\beta
\approx 1$) and a photon, the single-collision luminosity can be simplified to
\begin{equation}
\mathcal{L}_{sc} =
\frac{1}{2\pi\sqrt{\frac{\lambda\beta_0}{4\pi}+\beta_x\varepsilon_x}\sqrt{\frac{
\lambda\beta_0}{4\pi}+\beta_y\varepsilon_y}}.
\end{equation}
Thus, Eq.~(\ref{tot_flux_lumin}) can be rewritten in a simple form
\begin{equation}
 N_{tot} = N_e N_p \mathcal{L}_{sc} \overline{\sigma_{tot}},
\end{equation}
where $\overline{\sigma_{tot}}$ is the total Compton scattering cross section
averaged over the momenta of the incident electrons and photons. Neglecting the
energy spread of the electrons and photons, $\overline{\sigma_{tot}}$ can be
approximated by $\sigma_{tot}$ of Eq.~(\ref{tot_scat_cross}),
which can be further simplified to the \textit{classical Thomson cross section}
if the recoil effect is negligible.

If the beam-beam collision rate is $f_0$, the gamma-ray flux is given by
\begin{equation}
 \frac{\mathrm{d}N_{tot}}{\mathrm{d}t} = N_e N_p \mathcal{L}_{sc}
\overline{\sigma_{tot}} f_0.
\end{equation}

\subsection{Spatial and energy distributions: semi-analytical calculation}
To obtain the spatial and energy distributions of a Compton gamma-ray beam, the
differential cross section should be used instead of the total cross section in
Eq.~(\ref{tot_flux_integration}). In addition, two constraints need to be
imposed during the integration of
Eq.~(\ref{tot_flux_integration})~\cite{Vladimir,Park_thesis}.  

\begin{figure}
\centering
\includegraphics[width=\columnwidth]{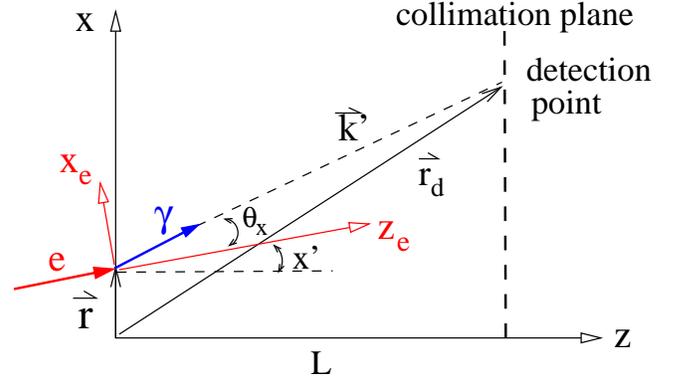}
\caption{\label{collimator-geometry}Geometric constraint for a scattered
gamma-ray photon. The diagram only shows the projection of the constraint in the
$x$-$z$ plane.}
\end{figure}

First, let us consider the geometric constraint, which assures the gamma-ray
photon generated
at the location $\vec{r}$ can reach the location $\vec{r}_d$ shown in
Fig.~\ref{collimator-geometry}. In terms of the position vector, this constraint
is given by
\begin{equation}
\frac{\vec{k}^\prime}{|\vec{k}^\prime|}=\frac{\vec{r}_d-\vec{r}}{|\vec{r}_d-\vec
{r}|},
\label{geometray_constraint}
\end{equation}
where $\vec{k}^\prime$ represents the momentum of the gamma-ray photon; $\vec{r}
= (x,y,z)$ denotes the location of the collision; and $\vec{r}_d=(x_d,y_d,z_d)$ denotes
the location where the scattered gamma-ray photon is detected.
Due to the finite spatial distribution and angular
divergence of the electron beam, a gamma-ray photon reaching the location
$\vec{r}_d$ can be scattered from an electron at different collision points with
different angular divergences. 

The constraint of Eq.~(\ref{geometray_constraint}) projected in the x-z and y-z
planes is given by
\begin{equation}
\theta_x+x^\prime = \frac{x_d-x}{L}, ~\theta_y+y^\prime = \frac{y_d-y}{L}.
\label{constraint_projected}
\end{equation}
Here, $\theta_x$ and $\theta_y$ are the projections of the scattering angle
$\theta_f$ in the x-z and y-z planes, i.e., $\theta_x = \theta_f\cos\phi_f$,
$\theta_y = \theta_f\sin\phi_f$ and $\theta_f^2 = \theta_x^2+\theta_y^2$, where
$\theta_f$ and $\phi_f$ are the angles defined in the electron coordinate system
($x_e,y_e,z_e$) in which the electron is incident along the $z_e$-direction
(Fig.~\ref{collimator-geometry}). $x^\prime$ and $y^\prime$ are the angular
divergences of the incident electron, i.e., the angles between the electron
momentum and $z$-axis. $L$ is the distance between the collision point and the
detection plane (or the collimation plane). Note that a far field detection (or
collimation) has been assumed, i.e.,  $L\gg |\vec{r}|$ and $ L\approx
|\vec{r}_d|$. 

The second constraint is the energy conservation. Due to the finite energy
spread of the electron beam, the gamma-ray photon with an energy of $E_g$ can be
produced by electrons with various energies and scattering angles. 
Mathematically, this constraint is given by 
\begin{equation}
\delta(\bar{E}_g-E_g),
\label{energycondition}
\end{equation}
where 
\begin{equation}
\bar{E}_g = \frac{4\bar{\gamma}^2E_p}{1+\bar{\gamma}^2\theta_f^2+4\bar{\gamma}
E_p/mc^2}.
\end{equation}

Imposing the geometric and energy constraints in
Eq.~(\ref{tot_flux_integration}), the spatial and energy distributions of a
Compton gamma-ray beam can be obtained by integrating all the individual
scattering events, i.e.,
\begin{eqnarray}
\frac{\mathrm{d}N(E_g,x_d,y_d)}{ \mathrm{d}\Omega_d
\mathrm{d}E_g}&\approx&N_eN_p\int
\frac{\mathrm{d}\sigma}{\mathrm{d}\Omega}\delta(\bar{E}
_g-E_g)c(1+\beta)\nonumber\\
&&\!\!\!\!\!\!\!\!\!\!\!\!\times f_e(x,y,z,x^\prime,y^\prime,p,t)\nonumber\\
&&\!\!\!\!\!\!\!\!\!\!\!\!\times
f_p(x,y,z,k,t)\mathrm{d}x^\prime\mathrm{d}y^\prime
\mathrm{d}p\mathrm{d}k\mathrm{d}V\mathrm{d}t,
\label{spatialenergydist}
\end{eqnarray}
where $\mathrm{d}\Omega_d = \mathrm{d}x_d \mathrm{d}y_d/L^2$, and
$\mathrm{d}\sigma/\mathrm{d}\Omega$ is the differential Compton scattering cross
section. Note that a head-on collision between electron and laser beams has been
assumed, and the density function $f_e(\vec{r},\vec{p},t)$ has been replaced
with $f_e(x,y,z,x^\prime,y^\prime,p,t)$ of Eq.~(\ref{electron-photon-dist})
under the approximation $p_z\approx p$ for a relativistic electron beam. In
addition, the integration $\int \cdots f_p(\vec{r},\vec{k},t)\mathrm{d}^3k$ is
replaced with $\int \cdots f_p(x,y,z,k,t) \mathrm{d}k$, where $f_p(x,y,z,k,t)$
is defined in Eq.~(\ref{electron-photon-dist}). Integrations over
$\mathrm{d}k_x$ and $\mathrm{d}k_y$ have been carried out since the differential
cross section has a very weak dependency on $k_x$ and $k_y$ for a relativistic
electron beam.    

Assuming head-on collisions for each individual scattering event ($\theta_i =
\pi$ and $\mathrm{d}\sigma/\mathrm{d}\Omega$ is given by
Eq.~(\ref{crosssection-1})), neglecting the angular divergences of the laser
beam  and replacing $x^\prime$ and $y^\prime$ with $\theta_x$ and $\theta_y$, we
can integrate Eq.~(\ref{spatialenergydist}) over $\mathrm{d}V,\mathrm{d}t$ and
$\mathrm{d}p$ to yield the following result (see Appendix~\ref{append}),
\begin{widetext}
\begin{eqnarray}
\frac{\mathrm{d}N(E_g,x_d,y_d)}{\mathrm{d}E_g \mathrm{d}x_d
\mathrm{d}y_d}&=&\frac{r_e^2L^2N_eN_p}{4\pi^3\hbar
c\beta_0\sigma_\gamma\sigma_k}\int^{\infty}_0
\int^{\sqrt{4E_p/E_g}}_{-\sqrt{4E_p/E_g}}\int^{\theta_{xmax}}_{-\theta_{xmax}}
\frac{1}{\sqrt{\zeta_x\zeta_y}\sigma_{\theta x}\sigma_{\theta
y}}\frac{\gamma}{1+2\gamma E_p/mc^2}\nonumber \\
&&\times\left\lbrace\frac{1}{4}\left[\frac{4\gamma^2E_p}{
E_g(1+\gamma^2\theta_f^2)}+\frac{E_g(1+\gamma^2\theta_f^2)}{4\gamma^2E_p}\right]
-2\cos^2(\tau-\phi_f)\frac{\gamma^2\theta_f^2}{(1+\gamma^2\theta_f^2)^2}
\right\rbrace \nonumber\\
&&\times\exp\left[-\frac{(\theta_x-x_d/L)^2}{2\sigma_{\theta_x}^2}-\frac{
(\theta_y-y_d/L)^2}{2\sigma_{\theta_y}^2}-\frac{(\gamma-\gamma_0)^2}{
2\sigma_\gamma^2}-\frac{(k-k_0)^2}{2\sigma_k^2}\right]\mathrm{d}\theta_x
\mathrm{d}\theta_y \mathrm{d}k,
\label{spatialenergydist3}
\end{eqnarray}
where
\begin{eqnarray}
\xi_x = 1+(\alpha_x-\frac{\beta_x}{L})^2+\frac{2k\beta_x\varepsilon_x}{\beta_0},
~\zeta_x =1+\frac{2k\beta_x\varepsilon_x}{\beta_0},~\sigma_{\theta x} =
\sqrt{\frac{\varepsilon_x\xi_x}{\beta_x\zeta_x}},\nonumber\\
\xi_y = 1+(\alpha_y-\frac{\beta_y}{L})^2+\frac{2k\beta_y\varepsilon_y}{\beta_0},
~\zeta_y =1+\frac{2k\beta_y\varepsilon_y}{\beta_0},~\sigma_{\theta y} =
\sqrt{\frac{\varepsilon_y\xi_y}{\beta_y\zeta_y}},\nonumber\\
\theta_f = \sqrt{\theta_x^2+\theta_y^2},~\theta_{xmax} =\sqrt{
4E_p/E_g-\theta_y^2},~\sigma_\gamma = \frac{\sigma_{E_e}}{mc^2},\nonumber\\
\gamma=\frac{2E_g
E_p/mc^2}{4E_p-E_g\theta_f^2}\left(1+\sqrt{1+\frac{4E_p-E_g\theta_f^2}{
4E_p^2E_g/(mc^2)^2}}\right), 
\end{eqnarray}
and $\sigma_{E_e}$ is the RMS energy spread of the electron beam. 

In a storage ring, the vertical emittance of the electron beam is typically much
smaller than the horizontal emittance. For a Compton scattering occurring at a
location with similar horizontal and vertical beta functions ($\beta_x \sim
\beta_y$), the vertical divergence of the electron beam can be neglected. In
addition, the photon energy spread of a laser beam is small, and its impact can
also be neglected in many practical cases. 
Under these circumstances, the cross section term in
Eq.~(\ref{spatialenergydist3}) has a weak dependence on $\theta_y$ ($\approx
y_d/L$) and $k$ ($\approx k_0$). With the assumption of an unpolarized or
circularly polarized laser beam, Eq.~(\ref{spatialenergydist3}) can be
simplified further after integrating $\theta_y$ and $k$: 

\begin{eqnarray}
\frac{\mathrm{d}N(E_g,x_d,y_d)}{\mathrm{d}E_g
\mathrm{d}x_d\mathrm{d}y_d}&\approx&
\frac{r_e^2 L^2N_eN_{p}}{2\pi^2 \hbar c
\beta_0\sqrt{\zeta_x}\sigma_\gamma\sigma_{\theta x}}
\int^{\theta_{xmax}}_{-\theta_{xmax}}\frac{\gamma}{1+2\gamma E_p/mc^2}
\left\lbrace \frac{1}{4}\left[\frac{4\gamma^2 E_p}{E_g(1+\gamma^2\theta_f^2)}+
\frac{E_g(1+\gamma^2\theta_f^2)}{4\gamma^2 E_p}\right]\right.\nonumber\\
&&\left.-\frac{\gamma^2\theta_f^2}{(1+\gamma^2\theta_f^2)^2}\right\rbrace
\times\exp\left[-\frac{(\theta_x-x_d/L)^2}{2\sigma_{\theta_x}^2}-
\frac{(\gamma-\gamma_0)^2}{2\sigma_\gamma^2}\right] \mathrm{d}\theta_x,
\label{dist_no_vertical}
\end{eqnarray}
\end{widetext}
where $\theta_{xmax}=\sqrt{4E_p/E_g-(y_d/L)^2}$. 

The integrations with respect to $k, ~\theta_y$ and $\theta_x$ in
Eq.~(\ref{spatialenergydist3}) or $\theta_x$ in Eq.~(\ref{dist_no_vertical})
must be carried out numerically. For this purpose, a numerical integration
Compton scattering code (CCSC) in the C++ computing language has been developed
to evaluate the integrals of Eqs.~(\ref{spatialenergydist3})
and~(\ref{dist_no_vertical}).

With the detailed spatial and energy distributions of the Compton gamma-ray beam
$\mathrm{d}N(E_g,x_d,y_d)/(\mathrm{d}E_g \mathrm{d}x_d\mathrm{d}y_d)$, the
energy spectrum of the gamma-ray beam collimated by a round aperture with a
radius of $R$ can be easily obtained by integrating
$\mathrm{d}N(E_g,x_d,y_d)/(\mathrm{d}E_g \mathrm{d}x_d\mathrm{d}y_d)$ over the
variables $x_d$ and $y_d$ for the entire opening aperture, i.e.,
$\sqrt{x_d^2+y_d^2}\leqslant R^2$. 

The transverse misalignment effect of the collimator on the gamma-ray beam
distributions can be introduced by replacing $x_d$ and $y_d$ with $x_d+\Delta x$ and
$y_d+\Delta y$ in Eq.~(\ref{spatialenergydist3}) or Eq.~(\ref{dist_no_vertical}),
where $\Delta x$ and $\Delta y$ are the collimator offset errors in the horizontal and
vertical directions, respectively.

\subsection{Spatial and energy distributions: Monte Carlo simulation}
In the previous section, we have derived an analytical formula to study the
spatial and energy distributions of a Compton gamma-ray beam. However, to
simplify the calculation several approximations have been made: head-on
collisions for each individual scattering event, a negligible angular divergence
of the laser beam, and far field collimation. 

A completely different approach to study Compton scattering process is to use a
Monte Carlo simulation. With this numerical technique, effects that cannot be easily included in
an analytical method can be properly accounted for. 
For example, using a Monte Carlo simulation we can study the scattering process
for an arbitrary collision angle. 
With this motivation, we developed a Monte Carlo Compton scattering code. In
following, the algorithm of this code is presented.

\subsubsection{Simulation setup}
At the beginning of the collision, both the electron and laser pulses are located some distance
away from the origin (Fig.~\ref{lab_laser_frame}), and two pulse centers
arrive at the origin at the same time ($t=0$). The collision duration is divided
into a number of time steps, and the time step number represents the time in the
simulation.

Due to a large number of electrons in the bunch, it is not practical to track
each electron in the simulation. Therefore, the electron bunch is divided into a
number of macro particles (for example, $10^6$) which are tracked
in the simulation. 

The phase space coordinates of each macro particle are sampled at time $t=0$.
For an electron beam with Gaussian distributions in phase space, the coordinates
are sampled according to the electron beam Twiss parameters as
follows~\cite{CAIN,chao_1} 
\begin{eqnarray}
 x(0) &=& \sqrt{2 u_1\varepsilon_x\beta_x }\cos\phi_1,\nonumber\\
x^\prime(0) &=&-\sqrt{2 u_1\varepsilon_x /\beta_x
}(\alpha_x\cos\phi_1+\sin\phi_1),\nonumber\\
 y(0) &=& \sqrt{2 u_2 \varepsilon_y \beta_y }\cos\phi_2,\nonumber\\
 y^\prime(0) &=&-\sqrt{2 u_2 \varepsilon_y
/\beta_y}(\alpha_y\cos\phi_2+\sin\phi_2),\nonumber\\
z(0) &=& \sigma_z r_1,\nonumber\\
 E_e&=&E_0(1+\sigma_{E_e} r_2),
\label{electron_parameters}
\end{eqnarray}
where $u_{1,2}$ are random numbers generated using an exponential distribution
with
a unit mean parameter (i.e., $e^{-u_{1,2}}$), $r_{1,2}$ are random numbers
generated according to a Gaussian distribution with a zero mean and unit
standard deviation, and $\phi_{1,2}$ are uniformly distributed random numbers
between $0$ and
$2\pi$. 
The coordinates of macro particles at any other time ($t\neq0$) can then be obtained by
transforming the coordinates given by Eq.~(\ref{electron_parameters}). 

The Compton scattering is simulated according to the local intensity and
momentum of the laser beam at the collision point. The intensity of the laser beam
at the collision point $(x,y,z)$ in the electron-beam coordinate system can be
calculated according to Eq.~(\ref{electron-photon-dist}) using the laser-beam
coordinates $(x_l,y_l,z_l)$ transformed from $(x,y,z)$. 
The momentum direction $\hat{k}$ of the photon at the collision point $(x,y,z)$
can be calculated from the point of view of electromagnetic wave of the photon
beam. For a Gaussian laser beam, its propagation phase $\psi(x_l,y_l,z_l)$ in
the laser-beam coordinate system is given by~\cite{CAIN,AESiegman} 
\begin{equation}
 \psi(x_l,y_l,z_l) =-ik_lz_l-ik_lz_l\frac{x_l^2+y_l^2}{2(\beta_0^2+z_l^2)};
\end{equation}
the wavevector (the momentum of photon $\vec{k}_l$) is given by $\vec{k}_l =
\bigtriangledown  \psi(x_l,y_l,z_l)$. Thus, 
\begin{equation}
\hat{k}_l\approx-\frac{1}{\sqrt{1+c_1^2+c_2^2}}(c_1\hat{x}_l+c_2\hat{y}_l+\hat{z
}_l),
\label{local_vector}
\end{equation}
where
\begin{equation}
 c_1 = \frac{x_l z_l}{\beta_0^2+z^2_l},~c_2 = \frac{y_l z_l}{\beta_0^2+z^2_l}.
\end{equation}
The unit vector $\hat{k}_l$ expressed in the electron-beam coordinate system
gives the momentum direction of the laser photon in this coordinate system.
\subsubsection{Simulation procedures}
At each time step, the Compton scattering process is simulated for each macro
particle. The simulation proceeds in two stages. In the first stage, the
scattering probability is calculated using the local intensity and momentum of
the laser beam. According to this probability, the scattering event is sampled.
If the scattering happens, a gamma-ray photon will be generated, and the
simulation proceeds to the next stage. In the second stage, the energy and
scattering angles (including the polar and azimuthal angles) of the gamma-ray
photon are sampled according to the
differential Compton scattering cross section. The detailed simulation
procedures for these two stages are presented as follows.

\subsubsection{First stage: scattering event}
Since the energy and scattering angles of the gamma-ray photon are not the
concern at this stage, the total scattering cross section is used to calculate
the scattering probability. 
According to Eq.~(\ref{collisions_rate}), the scattering probability
$P(\vec{r},\vec{p},\vec{k},t)$ in the time step $\Delta t$ for the macro
particle at the collision point $(x,y,z)$ is given by
\begin{equation}
P(\vec{r},\vec{p},\vec{k},t)
=\sigma_{tot}(\vec{p},\vec{k})
c(1-\vec{\beta}\cdot\vec{k}/|\vec{k}|)n_p(x,y,z,k,t)\Delta t,
\label{collisions_prob}
\end{equation}
where $n_p(x,y,z,k,t)$ and $\vec{k}$ are the local density and wavevector of the
photon beam, respectively; $\sigma_{tot}(\vec{p},\vec{k})$ is the total
scattering cross section given by Eq.~(\ref{tot_scat_cross}).

According to the probability $P(\vec{r},\vec{p},\vec{k},t)$, the scattering
event is sampled using the \textit{rejection} method as
follows~\cite{penelope,Nelson:1985ec}: first, a random number $r_3$ is uniformly
generated in the range from $0$ to $1$; if $r_3\leq P(\vec{r},\vec{p},\vec{k},t)$,
Compton scattering happens; otherwise the scattering does not happen, and the
above sampling process is repeated for the next macro particle. 

\subsubsection{Second stage: scattered photon energy and direction}

When a Compton scattering event happens, a gamma-ray photon is generated. The
simulation proceeds to the next stage to determine the energy and scattering
angles
 of the gamma-ray photon. For convenience, the sampling probability for
generating gamma-ray photon parameters is calculated in the electron-rest frame
coordinate system $(x_e^\prime,y_e^\prime,z_e^\prime)$ in which the electron is
at rest and the laser photon is propagated along the $z_e^\prime$-axis
direction. 

Since the momenta of macro particles and laser photons have been expressed in
the electron-beam coordinate system ($x,y,z$) in the lab frame, we need to
transform the momenta to those defined in the electron-rest frame coordinate
system $(x_e^\prime, y_e^\prime, z_e^\prime)$.  After transformations, the
sampling probability for generating the scattered gamma-ray photon energy and
direction will be calculated as follows.

In the electron-rest frame coordinate system $(x_e^\prime, y_e^\prime,
z_e^\prime)$, according to Eq.~(\ref{scatteredphotonenergy}) the scattered
photon energy is given by 
\begin{equation}
 \frac{1}{E_g^\prime} =
\frac{1}{E_p^\prime}+\frac{1}{mc^2}(1-\cos\theta^\prime),
\label{gamma_energy_rest_frame}
\end{equation}
where $\theta^\prime$ is the scattering angle between the momenta of the
scattered and incident photons; $E_g^\prime$ and $E_p^\prime$ are the energies
of the scattered and incident photons, and $E_g^\prime$ is in the range of
\begin{equation}
 \frac{E_p^\prime}{1+2E_p^\prime/mc^2}\leq E_g^\prime \leq E_p^\prime.
\label{omega_range}
\end{equation}

In the electron-rest frame coordinate system, we can simplify the Lorentz
invariant quantities $X$ and $Y$ of Eq.~(\ref{angular_dif_crosssection}) to 
$X = 2E_p^\prime/mc^2$ and $Y = 2E_g^\prime/mc^2$. As a result, the
differential cross section is given by 
\begin{eqnarray}
 \frac{\mathrm{d}^2\sigma}{\mathrm{d}E_g^\prime
\mathrm{d}\phi^\prime}\!\!\!&=&\!\!\!\frac{ mc^2 r_e^2}{2
E_p^{\prime2}}\!\!\left\lbrace 
\!\!\left[1+P_t
\cos(2\tau^\prime\!-\!2\phi^\prime)\right]\!\!\left[\!\!\left(\frac{mc^2}{
E_p^\prime}\!-\!\frac{mc^2}{
E_g^\prime}\right)^2\right. \right.\nonumber\\
\!\!\!&&\left.\left.+2\left(\frac{mc^2}{E_p^\prime}-\frac{mc^2}{E_g^\prime}
\right)\right]+\frac{E_p^\prime}{E_g^\prime}
+\frac{E_g^\prime}{E_p^\prime}\right\rbrace,
\label{cross_section}
\end{eqnarray}
where $\tau^\prime$ is the azimuthal angle of the linear polarization direction
of the
incident photon beam defined in the system $(x_e^\prime, y_e^\prime,
z_e^\prime)$, and $\phi^\prime$ is the azimuthal angle of the scattered
photon. Note that the quantity $P_t$, the degree of linear polarization of the
incident photon beam, is invariant under Lorentz transformations.

The scattered photon energy $E_g^\prime$ and the azimuthal angle $\phi^\prime$
are sampled according to the differential cross section
Eq.~(\ref{cross_section}). Since Eq.~(\ref{cross_section}) depends on both
$E_g^\prime$ and $\phi^\prime$, the \textit{composition and rejection} sampling
method~\cite{penelope,Nelson:1985ec} is used to sample these two variables. To
sample the scattered gamma-ray photon energy $E_g^\prime$,
Eq.~(\ref{cross_section}) needs to be summed over the azimuthal angle
$\phi^\prime$ and written as 
\begin{equation}
  \frac{\mathrm{d}\sigma}{\mathrm{d}E_g^\prime}=\pi
r_e^2\frac{mc^2}{E_p^{\prime2}}(2+\frac{2E_p^\prime}{mc^2})f(E_g^\prime),
\end{equation}
where 
\begin{eqnarray}
f(E_g^\prime) &=&
\frac{1}{2+2E_p^\prime/mc^2}\left[\left(\frac{mc^2}{E_p^\prime}-\frac{mc^2}{
E_g^\prime}\right)^2\right.\nonumber\\
&&+2\left(\frac{mc^2}{E_p^\prime}-\frac{mc^2}{E_g^\prime}\right) 
\left.+\frac{E_p^\prime}{E_g^\prime} +\frac{E_g^\prime}{E_p^\prime}\right],
\end{eqnarray}
and $0\leq f(E_g^\prime)\leq 1$ for any $E_g^\prime$.
Now, the scattered gamma-ray photon energy $E_g^\prime$ can be sampled according
to $f(E_g^\prime)$ as follows: first, a uniform random number $E_g^\prime$ is
generated in the range given by Eq.~(\ref{omega_range}), and $r_4$ in the range 
from $0$ to $1$; if $r_4\leq f(E_g^\prime)$, $E_g^\prime$ is accepted, otherwise the
above sampling process is repeated until $E_g^\prime$ is accepted. If
$E_g^\prime$ is accepted, the scattering angle $\theta^\prime$ can be calculated
using Eq.~(\ref{gamma_energy_rest_frame}).

After the scattered gamma-ray photon energy $E_g^\prime$ is determined, the
azimuthal $\phi^\prime$ angle is sampled according to 
\begin{equation}
g(\phi^\prime) =\frac{\mathrm{d}^2\sigma}{\mathrm{d}E_g^\prime
\mathrm{d}\phi^\prime} /\frac{\mathrm{d}\sigma}{\mathrm{d}E_g^\prime}.
\end{equation}

After obtaining the gamma-ray photon energy $E_g^\prime$, and the angles
$\theta^\prime$ and $\phi^\prime$ in the electron-rest frame coordinate system,
we need to transform these parameters to those in the lab-frame coordinate
system.
In the meantime, the momentum of the scattered electron is also computed. This
electron can still interact with the laser photon in following time steps, which
allows to correctly model the multiple scattering process between the electrons
and laser photons.

\section{Benchmark and applications of Compton scattering codes}
Based upon the algorithms discussed in Section III, we have developed two
computer codes using the C$++$ programming language: the numerical integration
Compton scattering code
CCSC
and the Monte Carlo Compton scattering code MCCMPT. 
Below, we briefly discuss the benchmark and applications of these two codes.

\subsection{Energy Distribution}

\begin{figure}
\centering
\begin{tabular}{c}
\includegraphics[width=\columnwidth]{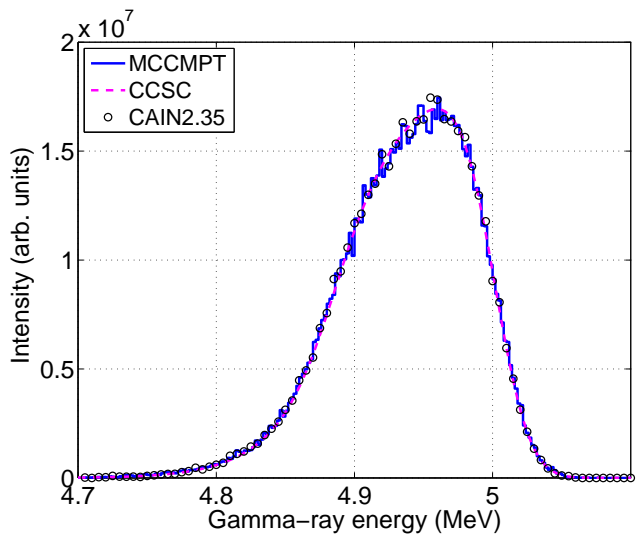}
\end{tabular}
\caption{\label{CCSC_MCCMPT} Compton gamma-ray beam energy spectra generated
using computer codes MCCMPT, CCSC and CAIN2.35. 
The stairs plot represents the spectrum simulated using the
code MCCMPT, the dash
line represents the spectrum calculated using the code CCSC, and the circles
represent the one
using the code CAIN2.35. The
electron beam energy and RMS energy spread are $400$~MeV and $0.2$\%,
respectively.
The electron beam  horizontal emittance is $10$~nm-rad, and the vertical
emittance is neglected. 
The laser wavelength is $600$~nm with negligible photon beam energy spread.
The gamma-ray beam is collimated by an aperture with a radius of $12$~mm located
$60$~meters
downstream from the collision point. }
\end{figure}

\begin{figure}
\centering
\begin{tabular}{c}
\includegraphics[width=\columnwidth]{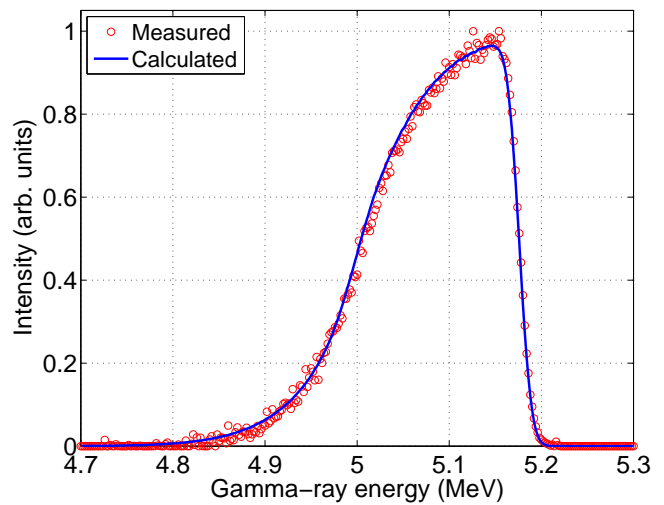}
\end{tabular}
\caption{\label{CCSC_Measured} Comparison between the measured and calculated
energy spectra of a Compton gamma-ray beam. The solid line represents the
calculated spectrum using the CCSC code, and the circles represent the measured
gamma-beam energy distribution after removing the escape peaks and Compton
plateau using a spectrum unfolding technique. The gamma-ray beam is
produced by Compton scattering of a $466$~MeV electron beam and a $790$~nm laser
beam at the HI$\gamma$S facility. 
The RMS energy spread of the electron beam is $0.1$\%, and horizontal and
vertical
emittance are $7.8$ and $1.0$~nm-rad, respectively.
The collimator with an aperture radius of $12.7$~mm is placed $60$~meters
downstream
from the collision point.}
\end{figure}

Our Compton scattering computer codes MCCMPT and CCSC have been benchmarked
against a well known
beam-beam colliding code CAIN2.35 developed at KEK for International Linear
Collider~\cite{CAIN}.
The energy spectra of Compton gamma-ray beams generated using these three codes
are shown in Fig.~\ref{CCSC_MCCMPT}.
We can see that these three codes can produce very close results. In terms of
computing time,
the codes CCSC,
MCCMPT and CAIN2.35 took about $10$~minutes, $150$~minutes and $1200$~minutes to
generate these
spectra using a single-core machine, respectively. Compared to the multi-purpose beam-beam
colliding code CAIN2.35, the dedicated Compton scattering codes CCSC and MCCMPT
are much faster and easy to use.

At the HI$\gamma$S facility, the Compton gamma-ray beam is usually measured
using a high-purity Germanium (HPGe) detector. Due to the non-ideal response of
the detector, the measured spectrum has a structure of a full energy peak, a
single and double escape peaks, and a Compton plateau. To unfold the measured
energy spectrum, a
novel end-to-end spectrum reconstruction method has been recently
developed~\cite{c.sun_2}. The comparison of the measured gamma spectrum and
calculated spectrum using the CCSC code
is shown in Fig.~\ref{CCSC_Measured}. A very good agreement between them is observed.

\begin{figure}
\centering
\begin{tabular}{c}
\includegraphics[width=\columnwidth]{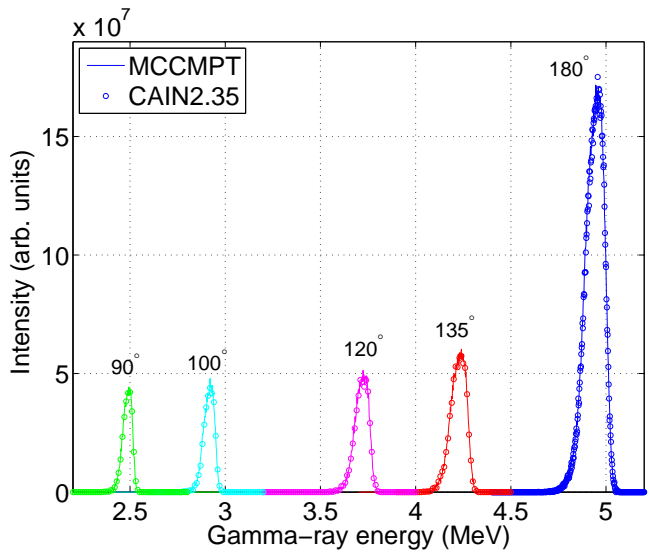}
\end{tabular}
\caption{\label{arb_ang} Compton gamma beam energy spectra for different
collision
angles $90^\circ$,~$100^\circ$,~$120^\circ$,~$135^\circ$ and ~$180^\circ$. These
spectra
are simulated using codes MCCMPT and CAIN2.35. The electron beam and
laser beam parameters are the same as those in Fig.~\ref{CCSC_MCCMPT}. The solid
lines represent
the spectra simulated using the code MCCMPT, and the circles represent the
spectrum simulated using the code
CAIN2.35.}
\end{figure}

Using the Monte Carlo simulation code, we can study the Compton scattering
process with
an arbitrary collision angle. The simulated spectra using MCCMPT are compared to
those using CAIN$2.35$ 
in Fig.~\ref{arb_ang}. Again, very good agreements are observed. It is clearly
shown
that the gamma-ray beam produced by a head-on collision of an electron and a
laser beams has the
highest energy and flux. With a 90 degree collision angle, the maximum energy of
the gamma-ray beam is only half of that for a head-on collision.

The energy spread of a Compton gamma-ray beam is mainly determined by the degree
of the collimation of the gamma beam, energy spread and angular divergence of
the electron beam~\cite{c.sun_2}. The contributions of these parameters to the
gamma-ray
beam energy spread are summarized in Table~\ref{depedences}. In some
literature~\cite{PhysRevE.54.5657,1983paac.conf...21S}, a simple quadratic sum
of individual contributions was used to estimate the energy spread of the
Compton scattering gamma-ray beam.
Since the electron beam angular divergence and the gamma-beam collimation introduce
non-Gaussian broadening effects on the gamma-beam spectrum~\cite{c.sun_2},
causing the spectrum to
have a long energy tail (Figs.~\ref{CCSC_MCCMPT} and~\ref{CCSC_Measured}), the
energy spread of the
gamma-ray beam cannot be given simply by the quadrature sum of different
broadening mechanisms. The realistic gamma-ray beam energy spread needs to be
calculated from its energy spectrum, which can be done using either the
numerical integration
code CCSC, or a Monte Carlo simulation code, MCCMPT or CAIN2.35.

\subsection{Spatial distribution}
\begin{figure*}
\centering
\begin{tabular}{cc}
\includegraphics[width=2in]{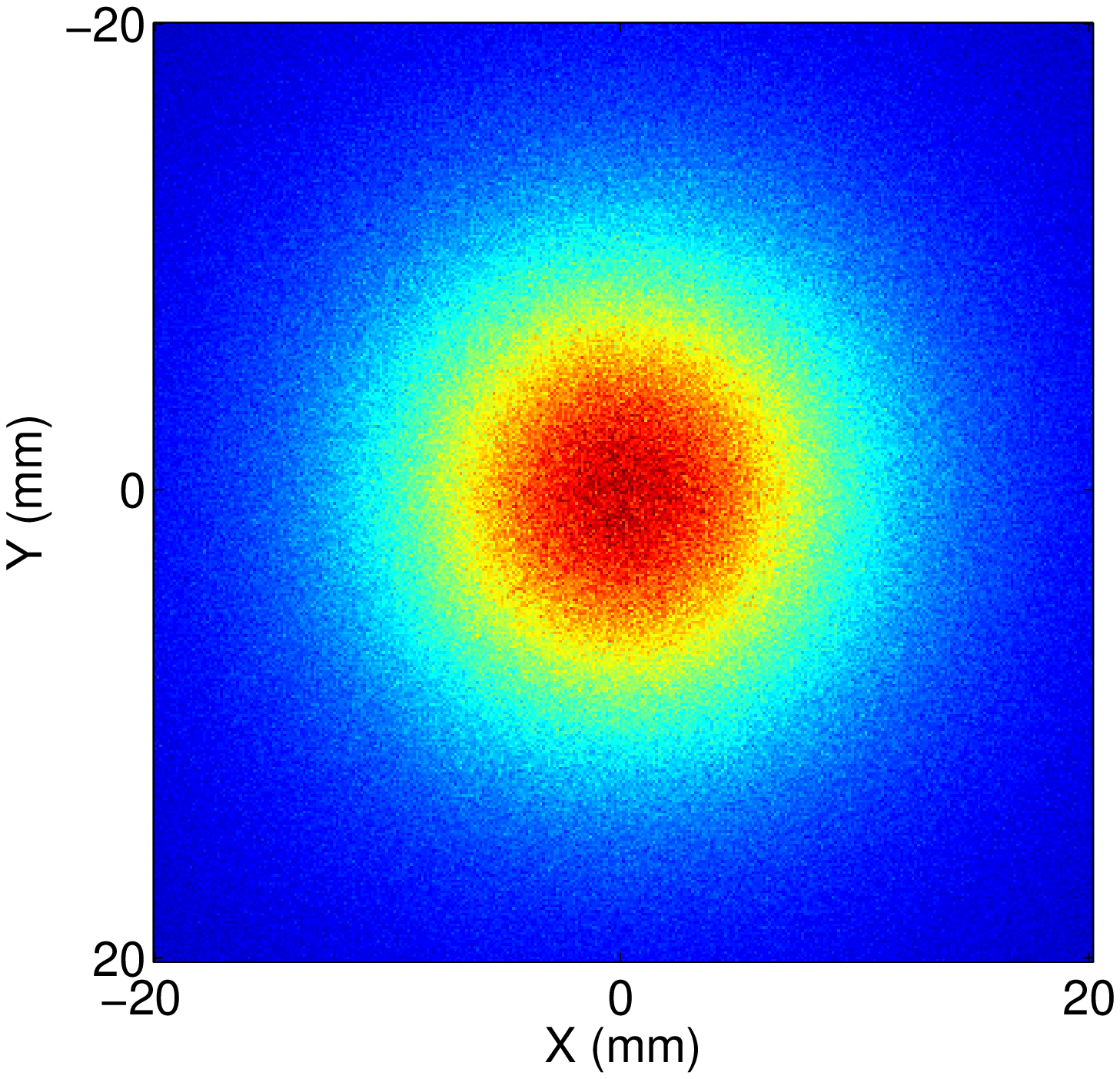}&
\includegraphics[width=2in]{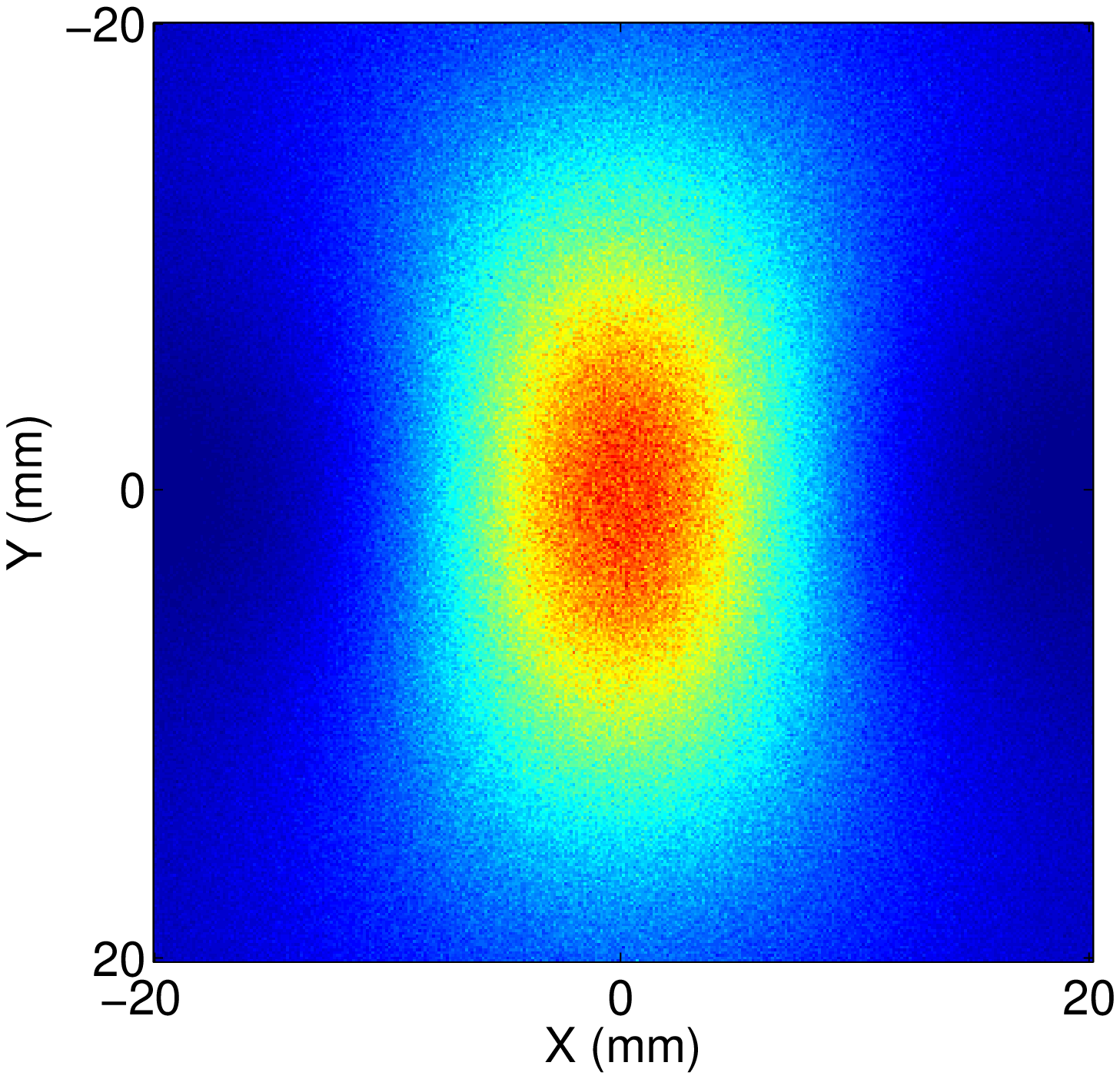} \\
\includegraphics[width=2in]{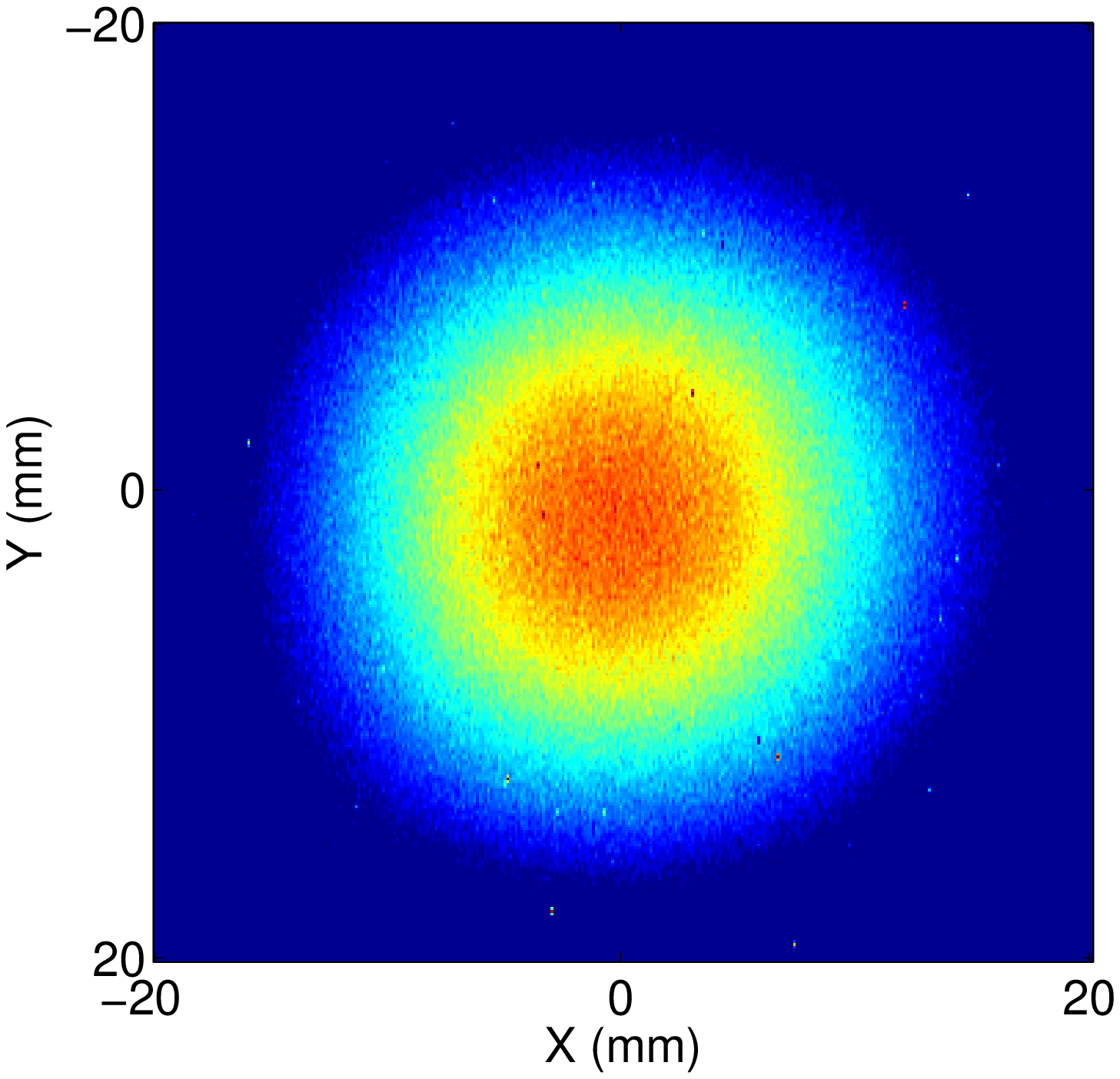} & 
\includegraphics[width=2in]{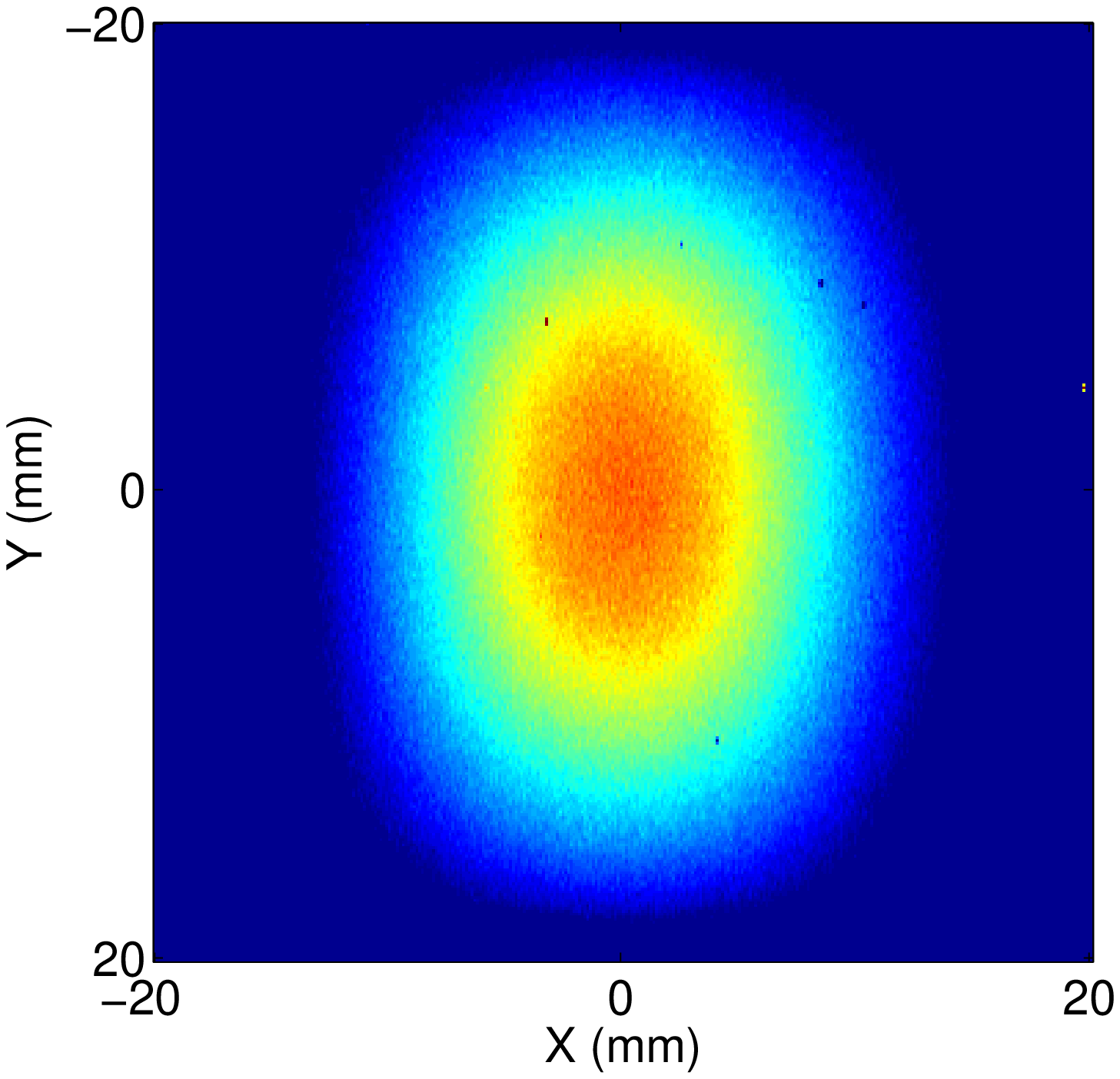} 
\end{tabular}
\caption{(Color) Spatial distributions of Compton gamma-ray beams for different
polarizations of the incoming laser beams. The gamma-ray beams are produced by
Compton scattering of a 680 MeV electron beam and a $378$~nm laser beam. The
observation plane is about $27$~meters downstream from the collision point. The
upper plots are the simulated images using the MCCMPT code. The lower ones are
the measured images.  The left images are produced using a circularly polarized
laser
beam; the right images with the linearly polarized laser beam.}
\label{simulated_measured_dist}
\end{figure*}

Figure~\ref{simulated_measured_dist} shows the spatial distribution of a
gamma-ray beam simulated by the MCCMPT code for circularly and linear polarized
incoming laser beams. For comparison, the measured spatial distributions of
gamma-ray beams using the recently developed gamma-ray imaging system at
HI$\gamma$S facility~\cite{c.sun_3} are also shown in
Fig.~\ref{simulated_measured_dist}. It can be seen that for a circularly
polarized incoming laser beam, the distribution is azimuthally symmetric; for a
linearly
polarized incoming laser beam, the gamma-ray beam distribution is asymmetric,
and is ``pinched'' along the direction of the laser beam polarization. 

More applications of using CCSC and MCCMPT codes to study characteristics of
Compton gamma-ray beams can be found in~\cite{c.sun_2,c.sun_1, c.sun_4}.

\section{\label{conclusion}Summary}

To study characteristics of a gamma-ray beam produced by Compton scattering of
an electron beam
and a laser beam, we have developed two algorithms: one based upon
an analytical calculation and the other using a Monte Carlo simulation.
According
to these algorithms, two computer codes, a numerical integration
code (CCSC) and a Monte Carlo simulation code (MCCMPT), have been developed at
Duke University.
These codes
have been extensively benchmarked against a beam-beam colliding code CAIN2.35
developed at
KEK and measurement results at the High Intensity
Gamma-ray Source (HI$\gamma$S) facility at Duke University. Using these two
codes, we are able to
characterize Compton gamma-ray beams with various
electron and laser beam parameters, arbitrary collision angles, and different
gamma-beam
collimation conditions. 

In this work, the nonlinear Compton scattering process is not considered, and
the polarization of the electron beam 
is not taken into account. Although the polarization of the gamma-ray
beam has been calculated in Section~II, this calculation is limited
to the particle-particle scattering case. Further studies will be carried out to
address these issues.

\begin{acknowledgments}
This work is supported in part by the US Department of Defense MFEL Program
as administered by the AROSR under contract number FA9550-04-01-0086 and by
the U.S. Department of Energy, Office of Nuclear Physics under grant number
DE-FG02-97ER41033. 
\end{acknowledgments}

\begin{widetext}
\appendix
\section{Spatial and energy distributions of a Compton gamma-ray
beam\label{append}}
The spatial and energy distributions of a Compton gamma-ray beam produced by a
head-on collision of an electron beam and a photon beam is given by 
\begin{equation}
\frac{\mathrm{d}N(E_g,x_d,y_d)}{\mathrm{d}\Omega_d \mathrm{d}E_g}\approx \int
\frac{\mathrm{d}\sigma}{\mathrm{d}\Omega}\delta(\bar{E}_g-E_g)c(1+\beta)n_e(x,y,
z,x^\prime,y^\prime,p,t)
n_p(x,y,z,k,t)\mathrm{d}x^\prime\mathrm{d}y^\prime\mathrm{d}p\mathrm{d}k\mathrm{
d}V\mathrm{d}t,
\label{spatialenergydist_append}
\end{equation}
where $\mathrm{d}\Omega_d = \mathrm{d}x_d \mathrm{d}y_d/L^2$;
$n_e(x,y,z,x^\prime,y^\prime,p, t)$ and $n_p(x,y,z,k,t)$ are the density
functions of the electron and photon beams given by Eq.
(\ref{electron-photon-dist});
$\mathrm{d}\sigma/\mathrm{d}\Omega$ is the differential cross section given by
Eq.~(\ref{crosssection-1}). For head-on collisions, we can simplify the
differential cross section to
\begin{equation}
\frac{\mathrm{d}\sigma}{\mathrm{d}\Omega}=8 r_e^2\left\lbrace
\frac{1}{4}\left[\frac{4\bar{\gamma}^2
E_p}{\bar{E_g}(1+\bar{\gamma}^2\theta_f^2)}+\frac{\bar{E_g}(1+\bar{\gamma}
^2\theta_f^2)}{4\bar{\gamma}^2
E_p}\right]-2\cos^2(\tau-\phi_f)\frac{\bar{\gamma}^2\theta_f^2}{(1+\bar{\gamma}
^2\theta_f^2)^2}\right\rbrace \left(\frac{\bar{E_g}}{4\bar{\gamma}
E_p}\right)^2.
\label{crosssection_headon_append}
\end{equation}

Replacing $x^\prime$ and $y^\prime$ with $\theta_x$ and $\theta_y$ according to
Eq.~(\ref{constraint_projected}), and neglecting the angular divergence of the laser
beam at the collision point, we can integrate
Eq.~(\ref{spatialenergydist_append}) over $\mathrm{d}V$ and $\mathrm{d}t$ and
obtain
\begin{eqnarray}
\frac{\mathrm{d}N(E_g,x_d,y_d)}{\mathrm{d}E_g \mathrm{d}x_d
\mathrm{d}y_d}&=&\frac{L^2N_e
N_p}{(2\pi)^3\beta_0\sigma_p\sigma_k}\int\frac{k}{\sqrt{\zeta_x\zeta_y}}\frac{1}
{\sigma_{\theta x}\sigma_{\theta_y}}
\frac{\mathrm{d}\sigma}{\mathrm{d}\Omega}\delta(\bar{E}
_g-E_g)(1+\beta)\nonumber\\
&&\times\exp\left[-\frac{(\theta_x-x_d/L)^2}{2\sigma_{\theta_x}^2}-\frac{
(\theta_y-y_d/L)^2}{2\sigma_{\theta_y}^2}-\frac{(p-p_0)^2}{2\sigma_p^2}-\frac{
(k-k_0)^2}{2\sigma_k^2}\right]\mathrm{d}\theta_x\mathrm{d}\theta_y\mathrm{d}
p\mathrm{d}k,
\label{spatialenergydist6}
\end{eqnarray}
where
\begin{eqnarray}
\xi_x = 1+(\alpha_x-\frac{\beta_x}{L})^2+\frac{2k\beta_x\varepsilon_x}{\beta_0},
~\zeta_x =1+\frac{2k\beta_x\varepsilon_x}{\beta_0},~\sigma_{\theta x} =
\sqrt{\frac{\varepsilon_x\xi_x}{\beta_x\zeta_x}}, \nonumber\\
\xi_y = 1+(\alpha_y-\frac{\beta_y}{L})^2+\frac{2k\beta_y\varepsilon_y}{\beta_0},
~\zeta_y =1+\frac{2k\beta_y\varepsilon_y}{\beta_0},~\sigma_{\theta y}
=\sqrt{\frac{\varepsilon_y\xi_y}{\beta_y\zeta_y}},\nonumber\\
\theta_f = \sqrt{\theta_x^2+\theta_y^2},~\theta_x =\theta_f\cos\phi_f,~\theta_y
=\theta_f\sin\phi_f.
\end{eqnarray}

Next, we need to integrate the electron beam momentum $\mathrm{d}p$. It is
convenient to change the momentum $p$ to the scaled electron beam energy
variable $\bar{\gamma}=E_e/(mc^2)$, and rewrite the delta-function
$\delta(\bar{E}_g-E_g)$ as 
\begin{equation}
\delta(\bar{E}_g-E_g)=\delta(\frac{4\bar{\gamma}^2E_p}{1+\bar{\gamma}
^2\theta_f^2+4\bar{\gamma}
E_p/mc^2}-E_g)=-\delta(\bar{\gamma}-\gamma)\frac{(1+\gamma^2\theta_f^2+4\gamma
E_p/mc^2)^2}{8\gamma E_p(1+2\gamma E_p/mc^2)},
\label{delta}
\end{equation}
where
\begin{equation}
\gamma=\frac{2E_g
E_p/mc^2}{4E_p-E_g\theta_f^2}\left(1+\sqrt{1+\frac{4E_p-E_g\theta_f^2}{
4E_p^2E_g/(mc^2)^2}}\right)
\end{equation}
is the root of
\begin{equation}
E_g=\frac{4\gamma^2E_p}{1+\gamma^2\theta_f^2+4\gamma E_p/mc^2}
\end{equation}
with the condition of $0\leq\theta_f\leq\sqrt{\frac{4E_p}{E_g}}$.

Substituting Eqs.~(\ref{crosssection_headon_append}),~(\ref{delta}) into Eq.
(\ref{spatialenergydist6}) and integrating $d\bar{\gamma}$, we can get
\begin{eqnarray}
\frac{\mathrm{d}N(E_g,x_d,y_d)}{\mathrm{d}E_g\mathrm{d}x_d\mathrm{d}y_d}&=&\frac
{r_e^2L^2N_eN_p}{4\pi^3\hbar c\beta_0\sigma_\gamma\sigma_k}\int^{\infty}_0
\int^{\sqrt{4E_p/E_g}}_{-\sqrt{4E_p/E_g}}\int^{\theta_{xmax}}_{-\theta_{xmax}}
\frac{1}{\sqrt{\zeta_x\zeta_y}\sigma_{\theta x}\sigma_{\theta
y}}\frac{\gamma}{1+2\gamma E_p/mc^2}\nonumber \\
&&\times\left\lbrace\frac{1}{4}\left[\frac{4\gamma^2E_p}{
E_g(1+\gamma^2\theta_f^2)}+\frac{E_g(1+\gamma^2\theta_f^2)}{4\gamma^2E_p}\right]
-2\cos^2(\tau-\phi_f)\frac{\gamma^2\theta_f^2}{(1+\gamma^2\theta_f^2)^2}
\right\rbrace \nonumber\\
&&\times\exp\left[-\frac{(\theta_x-x_d/L)^2}{2\sigma_{\theta_x}^2}-\frac{
(\theta_y-y_d/L)^2}{2\sigma_{\theta_y}^2}-\frac{(\gamma-\gamma_0)^2}{
2\sigma_\gamma^2}-\frac{(k-k_0)^2}{2\sigma_k^2}\right]\mathrm{d}\theta_x\mathrm{
d}\theta_y\mathrm{d}k,
\label{spatialenergydist3_append}
\end{eqnarray}
where 
\begin{equation}
\theta_{xmax}=\sqrt{4E_p/E_g-\theta_y^2}.
\end{equation}

\end{widetext}

\bibliography{compton_scattering}

\end{document}